\documentclass[11pt]{revtex4-2}
\usepackage{graphicx}
\usepackage{dcolumn}
\usepackage{bm}
\usepackage{amsmath}
\usepackage{amsfonts} 
\usepackage{amssymb}
\usepackage{latexsym}
\usepackage{bbm}
\usepackage{amssymb}
\usepackage{xcolor,ulem}
\usepackage{comment}
\usepackage{amsthm}
\usepackage{epsf}
\usepackage{epsfig}
\usepackage{caption3}
\usepackage{appendix}
\usepackage{cases}
\usepackage{subfigure}
\usepackage{multirow}
\usepackage{float}

\usepackage{booktabs}
\usepackage{threeparttable}
\usepackage{mathtools}
\DeclarePairedDelimiter{\abs}{\lvert}{\rvert}

\newcommand{\beq}{\begin {equation}}
\newcommand{\eeq}{\end   {equation}}
\newcommand{\bea}{\begin {eqnarray}}
\newcommand{\eea}{\end   {eqnarray}}
\newcommand{\baa}{\begin {array}   }
\newcommand{\eaa}{\end   {array}   }
\newcommand{\bit}{\begin {itemize} }
\newcommand{\eit}{\end   {itemize} }
\newcommand{\be }{\begin {equation}}
\newcommand{\ee }{\end   {equation}}

\begin{document}

\title{\normalsize Discovering a light charged Higgs boson via $W^{\pm *}$ + 4$b$ final states at the LHC}
\author{Z. Li}
\email[]{\small wangyan@imnu.edu.cn}
\affiliation{\small College of Physics and Electronic Information,
Inner Mongolia Normal University, Hohhot 010022, PR China}

\author{A. Arhrib }
\email[]{\small aarhrib@gmail.com}
\affiliation{\small Abdelmalek Essaadi University, Faculty of Sciences and Techniques,
B.P. 2117 Tétouan, Tanger, Morocco}

\author{R. Benbrik}
\email[]{\small r.benbrik@uca.ac.ma}
\affiliation{\small Laboratoire de Physique Fondamentale et Appliquée de Safi, Faculté Polydisciplinaire de Safi,
Sidi Bouzid, B.P. 4162, Safi, Morocco}

\author{M. Krab}
\email[]{\small mohamed.krab@usms.ac.ma}
\affiliation{\small Research Laboratory in Physics and Engineering Sciences, Modern and Applied Physics Team, Polydisciplinary Faculty, Beni Mellal, 23000, Morocco}

\author{B. Manaut}
\email[]{\small bmanaut@usms.ma}
\affiliation{\small Research Laboratory in Physics and Engineering Sciences, Modern and Applied Physics Team, Polydisciplinary Faculty, Beni Mellal, 23000, Morocco}

\author{S. Moretti}
\email[]{\small s.moretti@soton.ac.uk; stefano.moretti@physics.uu.se}
\affiliation{\small School of Physics and Astronomy, University of Southampton,
Southampton SO17 1BJ, UK\\
Department of Physics and Astronomy, Uppsala University, Box 516, SE-751 20 Uppsala, Sweden}

\author{Y. Wang}
\email[]{\small wangyan@imnu.edu.cn}
\affiliation{\small College of Physics and Electronic Information,
Inner Mongolia Normal University, Hohhot 010022, PR China \\
Inner Mongolia Key Laboratory for Physics and Chemistry of Functional Materials,
Inner Mongolia Normal University, Hohhot, 010022, China}

\author{Q.S. Yan}
\email[]{\small yanqishu@ucas.ac.cn}
\affiliation{\small Center for Future High Energy Physics, Chinese Academy of Sciences,
Beijing 100049, P.R. China \\ School of Physics Sciences, University of Chinese Academy of Sciences,
Beijing 100039, P.R. China}

\begin{abstract}
\end{abstract}
\maketitle
\newpage
\centerline{\bf Abstract}
\noindent
{\small
Most of the current experimental searches for charged Higgs bosons at the Large Hadron Collider (LHC) concentrate upon the $tb$ and $\tau\nu$ decay channels. In the present study, we analyze instead the feasibility of the bosonic decay channel $W^{\pm *} h$, with the charged gauge boson being off-shell and $h$ being a neutral light Higgs boson, which  decays predominantly into $b\bar{b}$.
We perform a Monte Carlo (MC) analysis for the associate production of a charged Higgs with such a light neutral one, $pp\to H^\pm h$, at the LHC followed by the aforementioned charged Higgs boson decay, which leads to a $W^{\pm *} +4b$ final state.
The analysis is performed within the  2-Higgs Doublet Model (2HDM) with Yukawa texture of Type-I.  We take into account all available experimental 
constraints from LEP, Tevatron and the LHC  as well as the theoretical requirements of self-consistency of this scenario.
In order to study the full process  $pp \rightarrow H^{\pm} h \rightarrow W^{\pm *} h h \rightarrow \ell^\pm \nu+ 4b$ ($\ell=e,\mu$), we provide several Benchmark Points (BPs) amenable to further analysis, with $M_{H^\pm}+M_{b} < M_{t}$, for which we prove that there is a strong  possibility that this spectacular signal could be found at the LHC with center of mass energy of 14 TeV and luminosity of 300 $\rm{fb}^{-1}$.
}

\section{Introduction}
The discovery of a $125$ GeV scalar particle at the Large Hadron Collider (LHC) \cite{Aad:2012tfa, Chatrchyan:2012ufa} represents the last piece of the Standard Model (SM). Generally speaking, the measured properties of this particle agree well with those predicted for the  SM Higgs boson ($H_{\rm SM}$) at the 2$\sigma$ level. However, there is still a possibility that the discovered scalar  belongs to an extended Higgs sector. Furthermore, most  new physics models with extra doublets (or triplets) consistently predict one or more charged Higgs bosons. Thus, if a charged Higgs boson is found at the LHC, it would be a clear evidence of new physics  with an extended Higgs sector structure.

We are well aware that the SM cannot be the ultimate theory of Nature and must only be an effective low energy theory of a more fundamental one originating at some high energy scale. Therefore, there must be other sectors in this fundamental theory that the SM does not account for and  that could explain some  limitation of it,  such as Dark Matter (DM), Charge and Parity (CP) violation, neutrino masses, etc. Leaving aside the fermionic (i.e., matter) and 
gauge (i.e., forces) sectors, we concentrate here on an extended Higgs sector. As Nature seems to privilege doublet representations, herein, we extend the SM Higgs sector by adding another doublet \cite{Lee:1973iz,Deshpande:1977rw,Branco:2011iw}. Such a Beyond the SM (BSM) scenario is known as the 2-Higgs Doublet Model (2HDM) (for a review, see, e.g.,  Ref.~\cite{Branco:2011iw}).

After Electro-Weak Symmetry Breaking (EWSB) takes place,  from the eight degrees of freedom initially present in the 2HDM,  three degrees of
freedom are used up as the longitudinal polarizations of the then massive $W^{\pm}$  and $Z$ bosons while
the remaining five ones become physical Higgs  particles, namely:  two CP-even (scalar) states ($h$ and
$H$ with $M_{h}<M_{H}$), a CP-odd (pseudoscalar) one ($A$) and two charged ones $H^{\pm}$. Herein, we assumed that the discovered Higgs state, $H_{\rm SM}$, coincides with the $H$ state of our 2HDM (the so-called inverted hierarchy scenario).  
In order to forbid  Flavor Changing Neutral Currents (FCNCs) at the tree level, a $Z_{2}$ symmetry is imposed, so  that each type of fermion only couples to one of the doublets in the 2HDM \cite{Glashow:1976nt}. Depending on the $Z_{2}$ charge assignments of the
Higgs doublets, there are four basic 2HDM (so-called) Types. In the Type-I case, in which we are interested here, all fermions
couple to a single Higgs doublet.

A charged Higgs boson can be produced and decayed at hadron colliders via a number of different processes (for a review, see, e.g., Ref.~\cite{Akeroyd:2016ymd}). In particular, the 
$pp \to t \bar{t}$ process can abundantly produce a light charged Higgs boson 
(with $M_{H^\pm} \leq m_t - m_b$) via $t\to b H^+$  decays (or the analogous antitop mode). Hence, for such a Higgs state, the production and decay mode
most often searched for is $pp \to t\bar{t}\to b\bar{b}H^- W^+$ + c.c., where the other top (anti)quark decays via the SM channel $t\to bW^+$.
In Ref.~\cite{Arhrib:2021xmc}, we showed that, for a light charged Higgs boson, its associated production with a  light neutral  Higgs state followed by the bosonic decays of the charged Higgs $H^\pm \to W^\pm h/A$  \cite{Akeroyd:1998dt,Arhrib:2016wpw}, may produce a number of $H^\pm$ bosons greater than the amount resulting from top (anti)quark decay. 
We emphasize that the production of $H^\pm$ through EW processes followed by $H^\pm \rightarrow W^\pm h/A$ has also been addressed in these works \cite{Bahl:2021str,Cheung:2022ndq,Mondal:2023wib,Bhatia:2022ugu,Bandyopadhyay:2015dio}.
In this note, we focus on the $pp \rightarrow W^{\pm *} \rightarrow H^{\pm} h \rightarrow W^{\pm*} h h \rightarrow l\nu b\bar bb\bar b$ process, wherein  $W^\pm$ is always off-shell and $h$ decays into $b\bar b$ pairs, by performing a full Monte Carlo (MC) analysis, include hard scattering, parton shower, hadronization and detector effects, for the emerging `$W^{\pm *}+4b$' final state. A similar signature arising from $pp \rightarrow H^{\pm} A \rightarrow W^{\pm} A A$ has recently been analysed in this work \cite{Sanyal:2023pfs}.  The main background is the $pp \rightarrow t\overline{t}$ process followed by SM top (anti)quark decays (henceforth, $t\bar t_{\ell\nu jj bb}$) while others include $W^{\pm*} b\bar bb\bar b$ (henceforth, $wbbbb$), $W^{\pm *}b\bar bjj$ (henceforth, $wjjbb$), $W^{\pm *}jjjj$ (henceforth, $wjjjj$) and $Zt\bar b$ + c.c. (henceforth, $ztb_{zjjbb}$), wherein $j$ represents a light quark or gluon jets and $b$ a $b$-jet. 

The  paper  is  organized  as  follows. In section~\ref{md},  we  briefly  discuss  the  2HDM  and  its Yukawa sector.  In section~\ref{psc}, we present the parameter space scans and discuss the applied constraints, finally giving six Benchmark Points (BPs). In section~\ref{cmc}, we perform a thorough collider analysis of such BPs and show how to establish the aforementioned signal for the 2HDM Type-I scenario. 
In section~\ref{sum}, we provide some conclusions.

\section{The 2HDM}\label{md}
The scalar sector of the 2HDM consists of two weak isospin doublets with hyper-charge $Y = 1$. The most general  Higgs  potential for the 2HDM that complies with the $\rm{SU(2)_L \times U(1)_Y}$ gauge structure of the EW sector has the following form  \cite{Branco:2011iw}:
\begin{eqnarray}
V(\phi_1,\phi_2) &=& m_{11}^2(\phi_1^\dagger\phi_1) +
m_{22}^2(\phi_2^\dagger\phi_2) -
[ m_{12}^2(\phi_1^\dagger\phi_2)+\text{h.c.}] ~\nonumber\\&& 
+ \frac12\lambda_1(\phi_1^\dagger\phi_1)^2 +
\frac12\lambda_2(\phi_2^\dagger\phi_2)^2 +
\lambda_3(\phi_1^\dag \lambda_4(\phi_1^\dagger\phi_2)(\phi_2^\dagger\phi_1) \nonumber  ~\nonumber\\ && +
\frac12\left[\lambda_5(\phi_1^\dagger\phi_2)^2 +\rm{h.c.}\right]
+~\left\{\left[\lambda_6(\phi_1^\dagger\phi_1)+\lambda_7(\phi_2^\dagger\phi_2)\right]
(\phi_1^\dagger\phi_2)+\rm{h.c.}\right\},
\label{CTHDMpot}
\end{eqnarray}
where $\phi_{1}$ and $\phi_2$ are the two Higgs doublet fields. By hermiticity of such a  potential,  $\lambda_{1,2,3,4}$ as well as $m_{11,22}^2$ are   real parameters while $\lambda_{5,6,7}$ and $m_{12}^2$  can be complex, in turn enabling possible Charge and Parity (CP) violation effects in the Higgs  sector.
Upon two minimization conditions of the potential, $m^2_{11}$ and $m^2_{22}$ can be replaced by $v_{1,2}$, which are the Vacuum Expectation Values (VEVs) of the Higgs doublets $\phi_{1,2}$, respectively. 
Moreover, the coupling $\lambda_{1,2,3,4,5}$ can be substituted by the four physical Higgs masses ($M_h, M_H, M_A$ and $M_{H^\pm}$) and the parameter $\sin(\beta-\alpha)$, where $\alpha$ and $\beta$ are, respectively, the mixing angles between CP-even  and CP-odd Higgs field components. Thus, the independent input  parameters are $M_{h}$, $M_{H}$, $M_{A}$, $M_{H^\pm}$, $\lambda_6$, $\lambda_7$, $\sin(\beta-\alpha)$, $\tan\beta$ and $m^2_{12}$.  

If both Higgs doublet fields of the general 2HDM couples to all fermions, the ensuing scenario can induce FCNCs in the Yukawa sector at tree level.  As intimated, to remedy this, a $Z_2$ symmetry is imposed  on the Lagrangian such that each fermion type interacts with only one of the Higgs doublets \cite{Glashow:1976nt}.  As a consequence, there are four possible types of 2HDM, namely Type-I, Type-II, Type-X (or lepton-specific) and Type-Y (or flipped). However, such a symmetry is explicitly broken by the quartic couplings $\lambda_{6,7}$ and softly broken by the (squared) mass term $m^2_{12}$. 
 In what follows, we shall consider a {CP}-conserving (i.e., $m^2_{12}$ and $\lambda_5$ are real) 2HDM Type-I and assume that  $\lambda_{6} = \lambda_{7} = 0$ to forbid the explicit breaking of $Z_2$,
while also taking $m^2_{12}$ to be generally small, thereby preventing  large FCNCs at tree level, which are   incompatible with experiment. 

In general, the couplings of the neutral and charged Higgs bosons to fermions can be described by the Yukawa Lagrangian  given by \cite{Branco:2011iw} 
\begin{eqnarray}
- {\mathcal{L}}_{\rm Yukawa} = \sum_{f=u,d,l} \left(\frac{m_f}{v} \kappa_f^h \bar{f} f h + 
\frac{m_f}{v}\kappa_f^H \bar{f} f H 
- i \frac{m_f}{v} \kappa_f^A \bar{f} \gamma_5 f A \right) + \nonumber \\
\left(\frac{V_{ud}}{\sqrt{2} v} \bar{u} (m_u \kappa_u^A P_L +
m_d \kappa_d^A P_R) d H^+ + \frac{ m_l \kappa_l^A}{\sqrt{2} v} \bar{\nu}_L l_R H^+ + \text{h.c.} \right),
\label{Yukawa-1}
\end{eqnarray}
where $\kappa_f^S$ ($S=h,H$ and $A$) are the Yukawa couplings in the 2HDM, which are illustrated 
in Tab.~\ref{yuk_coupl} for the Type-I under consideration. Here, $V_{ud}$ refers to a CKM matrix element and $P_{L,R}$ denote the left- and right-handed projection operators. The coupling of the two CP-even states $h$ and $H$ to gauge bosons $VV$ ($V = W^\pm, Z$) are proportional to $\sin(\beta-\alpha)$ and $\cos(\beta-\alpha)$, respectively. Since, if we assume that either $h$ or $H$ can be the observed SM-like Higgs boson, the coupling  to gauge bosons is obtained for $h$ when $\cos(\beta-\alpha) \rightarrow 0$ and for $H$ when $\sin(\beta-\alpha) \rightarrow 0$. 
Therefore, each scenario can explain the 125 GeV Higgs signal at the LHC. 
Following our works \cite{Arhrib:2021xmc,Wang:2021pxc,Arhrib:2021yqf,Wang:2021zjp}, though, we shall focus in the present paper on the scenario where $H$ mimics the observed signal with mass $\sim\,125$ GeV (as previously intimated).      

\begin{table}[H]
	\centering
	\renewcommand{\arraystretch}{1.2} %
	\setlength{\tabcolsep}{1.2pt}
	\begin{tabular}{c|c|c|c} 
		& $\kappa_u^{S}$ &  $\kappa_d^{S}$ &  $\kappa_\ell^{S}$  \\   \hline
		$h$~ 
		& ~ $  \cos\alpha/ \sin\beta$~
		& ~ $  \cos\alpha/ \sin\beta$~
		& ~ $  \cos\alpha/ \sin\beta $~ \\
		$H$~
		& ~ $  \sin\alpha/ \sin\beta$~
		& ~ $  \sin\alpha/ \sin\beta$~
		& ~ $  \sin\alpha/ \sin\beta$~ \\
		$A$~  
		& ~ $  \cot \beta $~  
		& ~ $  -\cot \beta $~  
		& ~ $  -\cot \beta $~  \\ 
	\end{tabular}
	\caption{Yukawa couplings of the fermions $f=u,d$ and $\ell$ to the neutral Higgs bosons $S=h,H$ and $A$ in the 2HDM Type-I.}
	\label{yuk_coupl}	
\end{table}	

\section{Parameter space scans and constraints}\label{psc}
With the goal to understand the 2HDM, a numerical exploration of the parameter space has been conducted in previous studies \cite{Arhrib:2021xmc,Arhrib:2021yqf} in order to identify regions of it that satisfy both theoretical requirements and experimental observations.
To facilitate this  process, the program \texttt{2HDMC-1.8.0} \cite{Eriksson:2009ws} was used. This publicly available software  allows systematic testing of parameter combinations under a wide range of theoretical and experimental constraints, as follows.

\begin{itemize}
	\item Vacuum stability constraints are enforced
in order to maintain the boundedness from below \cite{Deshpande:1977rw} of the 
  Higgs potential given in Eq. (\ref{CTHDMpot}). These constraints were implemented to ensure this requirement is of utmost importance, as it guarantees the stability of the vacuum state of the 2HDM. In other words, the vacuum stability constraints play a crucial role in preventing the potential from diverging to negative infinity, thereby ensuring the overall stability and reliability of the model construction. These constraints read as  
	\begin{align}
	\lambda_1 > 0,\quad\lambda_2>0, \quad\lambda_3>-(\lambda_1\lambda_2)^{1/2} ,\quad \lambda_3+\lambda_4-|\lambda_5|>-(\lambda_1\lambda_2)^{1/2}.
	\end{align}

	\item Perturbativity constraints were also taken into account during the analysis. These constraints impose limits on the quartic couplings of the  Higgs potential, by requiring that the absolute values of these couplings, denoted as $\lambda_i$ ($i=1,...,5$), satisfy $\abs{\lambda_i} \leq 4\pi$ \cite{Chang:2015goa}. Adhering to these perturbativity constraints ensures that the interactions in the model remain within a perturbative regime, where the calculated results remain reliable and valid. 
	
	\item Tree-level perturbative unitarity constraints play a vital role in ensuring the validity and consistency of the model scattering amplitudes at high energies. These constraints enforce that the amplitudes of various scattering processes involving (pseudo)scalars, vectors and (pseudo)scalar-vector interactions remain unitary. To satisfy these constraints, the absolute values of the following quantities must be limited to be less than $8\pi$ \cite{Kanemura:1993hm, Akeroyd:2000wc}:
	\begin{equation}
	\abs{a_{\pm}}, \abs{b_{\pm}}, \abs{c_{\pm}}, \abs{f_{\pm}}, \abs{e_{1,2}}, \abs{f_1}, \abs{p_1} < 8\pi,
	\end{equation}
	where  
	\begin{eqnarray}
	&&
	a_{\pm} = \frac{3}{2}(\lambda_1+\lambda_2)\pm\sqrt{\frac{9}{4}(\lambda_1-\lambda_2)^2+(2\lambda_3+\lambda_4)^2},
	\nonumber \\
	&&
	b_{\pm}=\frac{1}{2}(\lambda_1+\lambda_2)\pm \frac{1}{2}\sqrt{(\lambda_1-\lambda_2)^2+4\lambda_4^2},  \nonumber \\
	&&
	c_{\pm}=\frac{1}{2}(\lambda_1+\lambda_2)\pm \frac{1}{2}\sqrt{(\lambda_1-\lambda_2)^2+4\lambda_5^2},  \nonumber \\
	&&e_1\,=\lambda_3+2\lambda_4 - 3 \lambda_5, \quad e_2 =\lambda_3 - \lambda_5, \quad p_1 =\lambda_3 - \lambda_5, \nonumber\\
	&&f_+=\lambda_3+2\lambda_4 + 3 \lambda_5, \quad f_- =\lambda_3 + \lambda_5, \quad f_1=\lambda_3 + \lambda_4.
	\label{eq:PertBounds}
	\end{eqnarray}
	
\item
	The parameter space exploration also considers the EW oblique parameters, denoted as $S$ and $T$ \cite{Peskin:1990zt, Peskin:1991sw}. These parameters are utilized to control the mass splitting between the Higgs states. To ensure consistency with experimental measurements \cite{Haller:2018nnx}, the following constraints are imposed:
\begin{align}
S = 0.04 \pm 0.08, \quad T = 0.08 \pm 0.07.
\end{align}
In order to assess their consistency at a $95\%$ Confidence Level (CL), the correlation factor between $S$ and $T$, which is $0.92$, is also taken into account. 
	
	\item To account for potential additional Higgs bosons, exclusion bounds at a $95\%$ CL  are enforced using the \texttt{HiggsBounds-5.9.0} program \cite{Bechtle:2020pkv}. This program systematically checks each parameter point against the $95\%$ CL exclusion limits derived from Higgs boson searches conducted at LEP, Tevatron and LHC experiments. 

	\item To ensure agreement with the measurements of the SM-like Higgs state, constraints are enforced using the \texttt{HiggsSignals-2.6.0} program \cite{Bechtle:2020uwn}. This program incorporates the combined measurements of the SM-like Higgs boson from  LHC Run-1 and Run-2. 
	
	\item Constraints from flavor physics are incorporated by utilizing the following observables : 
	\begin{itemize}
		\item BR$(B \to X_s \gamma) = (3.32 \pm 0.15) \times 10^{-4}$ \cite{HFLAV:2016hnz},
		\item BR$(B_s \to \mu^+ \mu^-) = (3.0 \pm 0.6 \pm 0.25) \times 10^{-9}$ \cite{LHCb:2017rmj},
		\item BR$(B \to \tau \nu) = (1.06 \pm 0.19) \times 10^{-4}$ \cite{HFLAV:2016hnz}. 
	\end{itemize}
	To compute these observables, the code \texttt{SuperIso v4.1} \cite{Mahmoudi:2008tp} is employed. 
\end{itemize}

To test the allowed parts of the  parameter space, we propose the six BPs given in Tab. \ref{t:bp}. As one can see from such a table, the charged Higgs boson is light since its mass varies from $85.50$ to $115.66$ GeV, so it can be produced in top (anti)quark decays. Also, in this set of BPs, the mass of the  neutral Higgs $h$ is always smaller than the $H^\pm$ mass so that decays of the $H^\pm$ state into $W^\pm h$ pairs are possible. However, the $W^\pm$ boson emerging from the decay will be off-shell since the mass separation between $H^\pm$ and $h$ is always smaller than the $W^\pm$ mass, i.e., $M_{H^\pm}-M_h < M_{W^\pm}$. Therefore, the charged lepton arising from it might be soft in all BPs. This is of relevance, for a twofold reason: on the one hand,  as we are 
 focusing here on  charged Higgs boson production in association with a light neutral one, i.e., $pp \rightarrow H^\pm h$,  its  cross section does not  reach the pb level and one should thus aim at minimizing losses exploiting the lepton kinematics; on the other hand,  given our decay signature,  $ H^\pm h \rightarrow W^{\pm *} h h \rightarrow \ell^\pm \nu + 4b$ ($\ell=e,\mu$), 
the lepton is the object used for triggering purposes, so its kinematics is bound to comply with the trigger requirements. 
\begin{table}[H]
	\begin{center}
		\begin{small}
			\begin{tabular}{|c| c| c| c| c| c| c| c| c| c|}
				\hline
				 & $M_{h}$ & $M_{H}$ & $M_{A}$ & $M_{H^\pm}$ & $\sin{(\beta-\alpha)}$ & $\tan{\beta}$ & $m_{12}$ & $\sigma(W^{\pm *}+4b)$ (fb)  \\
				\hline
				BP1 & 65.11 & 125.00 & 112.07 &88.51 & $-0.061$ & 51.14 & 82.33 & 807.69 \\
				\hline
				BP2 & 69.88 & 125.00 & 108.31 & 85.50 & $-0.059$ & 41.90 & 113.63 & 675.55 \\
				\hline
				BP3 & 69.12 & 125.00 & 106.14 & 90.62 & $-0.092$ & 40.63 & 115.73 & 664.89 \\
				\hline
				BP4 & 64.39 & 125.00 & 107.74 &107.61 & $-0.059$ & 45.03 & 90.47 & 521.93 \\
				\hline
				BP5 & 65.20 & 125.00 & 104.30& 106.02 & $-0.064$ & 57.64 & 73.50 & 525.88 \\
				\hline
				BP6 & 68.65 & 125.00 & 114.53 & 115.66 & $-0.098$ & 48.67 & 96.16 & 397.13 \\
				\hline
			\end{tabular}
			\caption{Input parameters and  Leading Order (LO) signal cross sections at the parton level (using 
				$\sqrt{s} = 14$ TeV) for each BP are presented. All masses are in GeV.}\label{t:bp}
		\end{small}
	\end{center}
\end{table} 
In short, these BPs provide valuable theoretical scenarios to further investigations of the 2HDM Type-I framework as well as challenging configurations for actual  experimental analysis. Their selection is guided by their ability to satisfy the various constraints, while also potentially producing observable signals at the LHC, making these promising targets of future phenomenological studies.

  \begin{figure}[H]
	\begin{center} 
		\includegraphics[height=5cm]{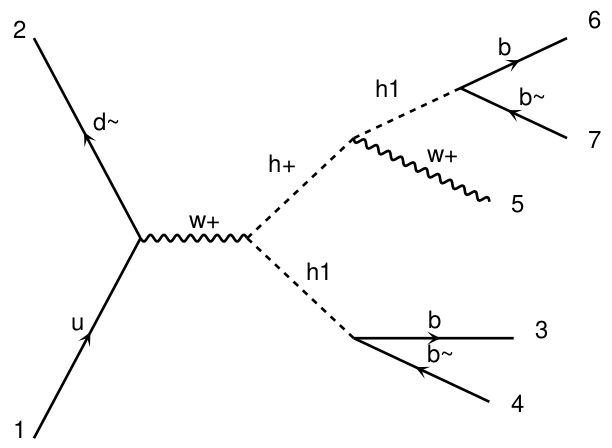}	
	\end{center}
	\caption{Feynman diagram for the signal (as generated by MadGraph). The symbol h1 denotes $h$.}
	\label{fig:FD}
\end{figure} 

\section{MC Analysis }\label{cmc}
In order to accurately analyze the signal (see Fig.~\ref{fig:FD}) and background events,  a comprehensive MC simulation is performed, accounting for  hard scattering as well as parton shower, hadronization and detector effects.  
\begin{itemize}
\item To compute the cross sections and generate events at the parton level for both signal and backgrounds, we utilize the \texttt{MadGraph5$\_$aMC@NLO-3.1.1} \cite{Alwall:2014hca} event generator. We adopt such a tool with default settings and a choice of Parton Distribution Functions (PDFs), including that of the factorization/renormalization scale. The normalisation of both signal and backgrounds is to the LO (for the signal, the inclusive cross section values are found in Tab.~\ref{t:bp}, as mentioned). 

	\item  Once the signal and background events are  generated at the parton level, we proceed to simulate the subsequent stages of particle interactions and decays using \texttt{Pythia-8.2} \cite{Sjostrand:2006za,Sjostrand:2014zea}. During the simulation in \texttt{Pythia}, the partons undergo showering processes, where additional gluons and quarks are emitted. The emitted partons subsequently hadronize, forming color-neutral hadrons such as mesons and baryons. Additionally, heavy flavor particles, e.g., charm and bottom hadrons, may decay into lighter particles.

	\item At the detector level, we use \texttt{Delphes-3.5.0} \cite{deFavereau:2013fsa} and, e.g.,  the default ATLAS card. We further adopt the  anti-$k_{t}$ jet algorithm
via FastJet   to cluster the final state partons  into jets. The choice of the jet parameter $\Delta R$ is important and we consider two values: 0.4 and 0.5. This parameter determines the size of the jets and affects their reconstruction and identification. To account for the mistagging of jets, we consider the efficiencies for $b$-jets, $c$-jets as well as light-quark and gluon  jets. The $b$-tagging efficiency is about $60\%$ to $70\%$, depending on the jet transverse momentum. The mis-tagging efficiency for a $b$-jet, which refers to the probability of a non-$b$-jet being misidentified as a $b$-jet, is approximately $0.2\%$ for a light quark and gluon jets while is around $10\%-14\%$ for a $c$-jet, again, dependent upon the jet transverse momentum.
 
\end{itemize}

\subsection{Acceptance cuts}
We start our analysis by applying acceptance cuts, which are imposed on variables such as pseudorapidity ($\eta$), transverse momentum ($p_T$), cone separation ($\Delta R$) and Missing $E_T$ (MET) to select the most relevant events for further analysis. The two sets of cuts, denoted as LACs (Loose Acceptance Cuts) and TACs (Tight Acceptance Cuts), are as follows:
\begin{align}
{\rm LACs:}\quad\quad
|\eta(\ell,{j})| < 2.5, \
p_T({j},\ell) > 10~\text{GeV}, \
\Delta R(\ell\ell/{jj}) > 0.4, \
\text{MET} > 5~\text{GeV},
\label{pc1}
\end{align}
\vspace*{-1.0truecm}
\begin{align}
{\rm TACs:}\quad\quad
|\eta(\ell,{j})| < 2.5, \
p_T({j},l) > 20~\text{GeV}, \
\Delta R(\ell\ell/{jj}) > 0.5, \
\text{MET} > 5~\text{GeV}.
\label{pc2}
\end{align}
In Tab.~\ref{t:parton_cross_section}, we tabulate the cross sections of  signal and background processes after these acceptance cuts. One can  observe that the signals are about 20-30 fb with LACs while they are 3-8 fb with TACs. 

  \begin{table}[H]
  \begin{center}
 \begin{small}
\begin{tabular}{|c|c|c|c|c|c|c||c|c|c|c|c|} 
\hline
$\sigma$ (fb) &  BP1&  BP2 &BP3 &BP4 &BP5 &BP6 & $t\bar{t}_{\ell\nu jj bb}$& $wbbbb$  & $wjjbb$&  $wjjjj$ & $ztb_{zjjbb}$\\
\hline
LACs & 32.59 & 20.93& 26.22 &31.94 &31.38 & 26.40 & 85625&   9.45  & 13474   & 789960 & 0.143\\
\hline
TACs & 5.39 & 2.71 &4.34 &8.31 &8.00  &7.89  & 54975  & 1.48 &2940 &  127545 & 9.3$\times 10^{-2}$ \\
  \hline
\end{tabular}
 \end{small}
  \caption{The cross sections of  signal (for our six BPs) and all background processes after the acceptance   cuts mentioned in the text.}\label{t:parton_cross_section}
 \end{center}
  \end{table}

\subsection{Pre-selection cuts}  
In order to reduce the number of background events, we have to resort to efficient $b$-tagging. For this purpose, we divide signal and background events in terms of the number of tagged $b$-jets, by defining three (multiplicity) categories:
\begin{itemize}
\item 4b0j:  four $b$-jets, no normal jets.
\item 3b1j:  three $b$-jets, one normal jet.
\item 2b2j:  two $b$-jets, two normal jets.
\end{itemize}  

Upon investigating Tab.~\ref{t:cross_section_bp23}, which contains the response to these cuts for both signal and background events in the three (multiplicity) categories identified, it 
 is noteworthy that the cross sections are rather small in general, which is due to the fact that lepton reconstruction and $b$-tagging efficiencies are dependent on the transverse momenta of the objects concerned (which are different for different BPs). As previously discussed, when the difference between $M_{H^\pm}$ and $M_{h}$ is small, the lepton will be soft and, when $M_{h}$ is small, the $b$-jets will be soft. In the end, these soft objects find it difficult to pass the TACs, so the rates are much larger for the LACs.  
  \begin{table}[H]
  \begin{center}
 \begin{small}
\begin{tabular}{|c|c|c|c|c|c|c||c|c|c|c|c|c|} 
\hline
BPs & BP1 & BP2 &BP3 &BP4 & BP5 & BP6   & $t\bar{t}_{\ell\nu jjbb}$& $wbbbb$  & $wjjbb$&  $wjjjj$ & $ztb_{zjjbb}$\\
\hline
LACs 4b0j & 1.39 & 0.86 & 1.16 & 1.78 & 1.74 & 1.67& 572.64  & 0.42 & 36.69   & 108.34 & 0.022\\
\hline
LACs 3b1j & 5.18 & 3.03 & 4.20 & 6.34 & 6.18& 5.72& 5226.43  & 1.51 & 354.22  & 699.25 & 0.054\\
\hline
LACs 2b2j & 8.28 & 4.71 & 6.64 & 10.22 & 9.83 & 9.03& 29583.0 & 2.67 & 2316.04 & 6480.41 & 0.073\\
\hline
\hline
TACs 4b0j & 0.15 & 0.08 & 0.13 & 0.31 & 0.31 & 0.34 & 98.96   & 8.6$\times 10^{-2}$  &  4.54  & 6.96   & 9.53$\times 10^{-3}$\\
\hline
TACs 3b1j & 0.47 & 0.21 & 0.38 & 1.01 & 0.95 & 0.99& 1658.4  & 2.61$\times 10^{-1}$ &  56.92 & 89.81  & 2.56$\times 10^{-2}$ \\
\hline
TACs 2b2j & 0.57 & 0.26 & 0.47 & 1.28 & 1.21 & 1.26 & 14704.8 & 3.34$\times 10^{-1}$ & 522.13 & 939.82 & 3.02$\times 10^{-2}$\\
\hline
\end{tabular}
 \end{small}
  \caption{The cross sections of  signal (for our six BPs) and all background processes after the pre-selection   cuts mentioned in the text.}
 \label{t:cross_section_bp23}
\end{center}
  \end{table}

\subsection{Kinematic observables for signal from background distinction}
In this subsection, we will reconstruct the resonances starting from the various final states. We take the {\rm 4b0j} category as an example. For the other two, light jets are treated as $b$-jets during all reconstructions.

In order to further improve the  signal-to-background ratio, we reconstruct the masses of the light Higgs boson, charged Higgs boson and charged gauge boson, so as to favor the signal. Simultaneously, in order to suppress top (anti)quark events (which are the dominant background), we also reconstruct the top  (anti)quark masses and veto these.

\begin{figure}[ht]
\begin{minipage}{0.47\textwidth}
      \begin{center} 
                     (a)	\\
         \includegraphics[height=4.5cm]{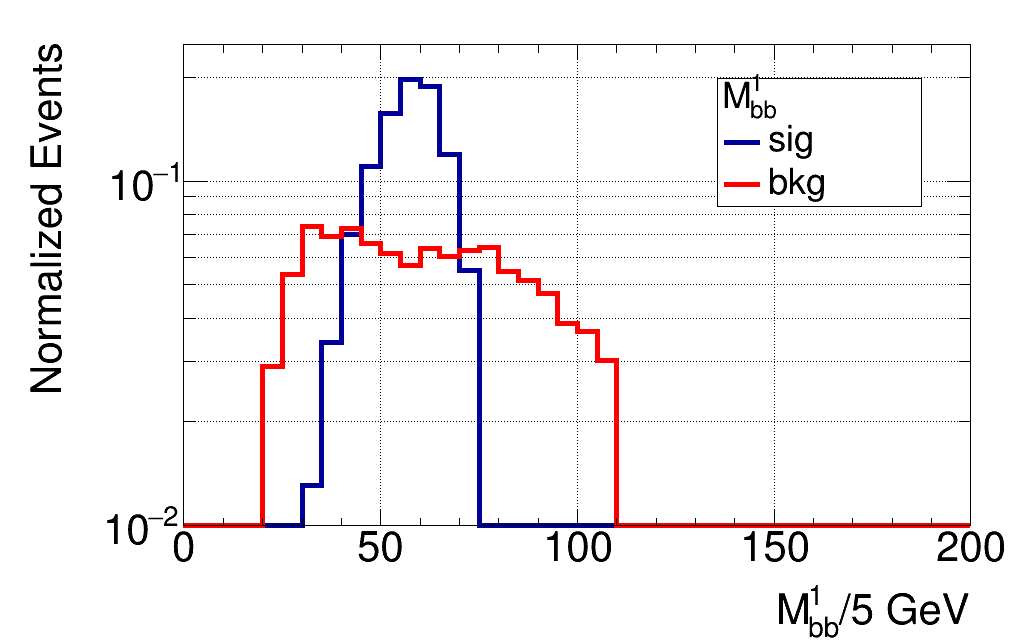}	
      \end{center}
\end{minipage}
\begin{minipage}{0.47\textwidth}
      \begin{center} 
                     (b)	\\
         \includegraphics[height=4.5cm]{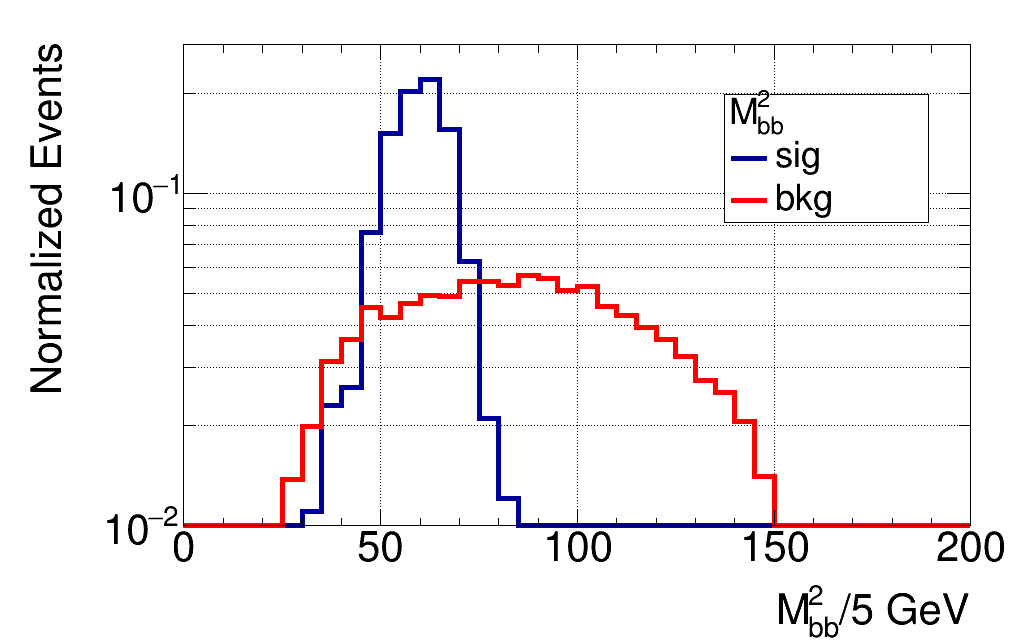}	
      \end{center}
\end{minipage}
 \caption{The light Higgs boson mass distributions $M_{b\bar{b}}^{1}$ and $M_{b\bar{b}}^{2}$ for BP4 and background events are shown.}\label{f:light_higgs}
\end{figure}

For signal events, we first look for four $b$-jets  to reconstruct two light Higgs bosons and find a two-by-two combination of them by minimizing the following $\chi$ square function:
\begin{equation}
\chi^{2}=(M^1_{b\bar b}-M_{h^{1}})^2+(M^2_{b\bar b}-M_{h^{2}})^2.
\end{equation}
Then we assign the reconstructed first light Higgs boson, $h^1$, to (say) the decay of the charged Higgs boson while the second,  $h^2$, is the light Higgs boson  produced in association with the charged Higgs boson. In Fig.~\ref{f:light_higgs}, we show the mass distributions of  the two light Higgs bosons for BP4 as an example. Hence, such a method can reconstruct the two light neutral Higgs bosons.

\begin{figure}[H]
	\begin{minipage}{0.47\textwidth}
		\begin{center} 
			(a)	\\
			\includegraphics[height=4.5cm]{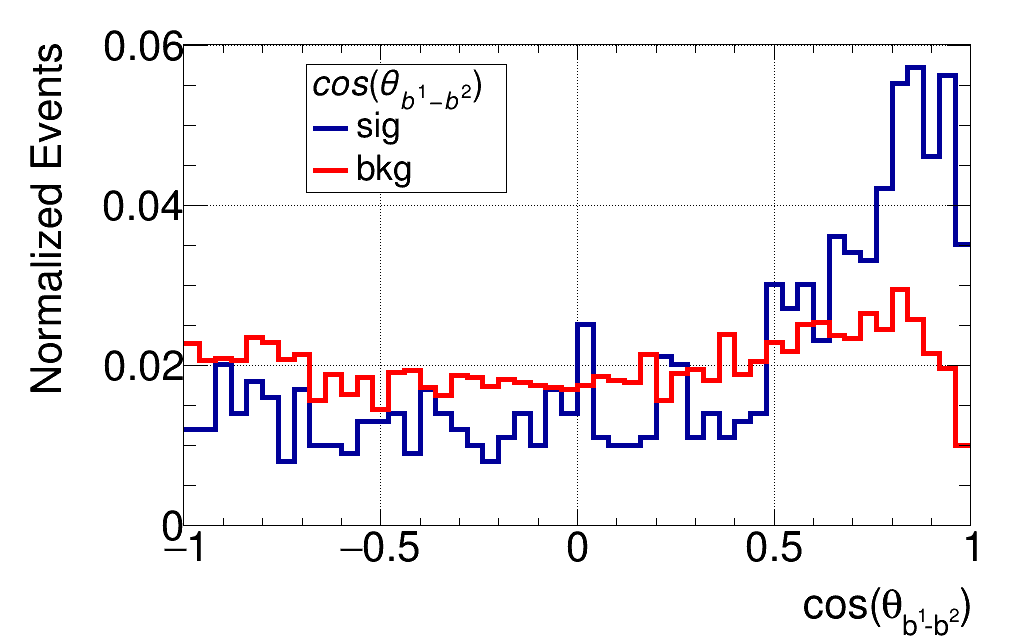}	
		\end{center}
	\end{minipage}
	\begin{minipage}{0.47\textwidth}
		\begin{center} 
			(b)	\\
			\includegraphics[height=4.5cm]{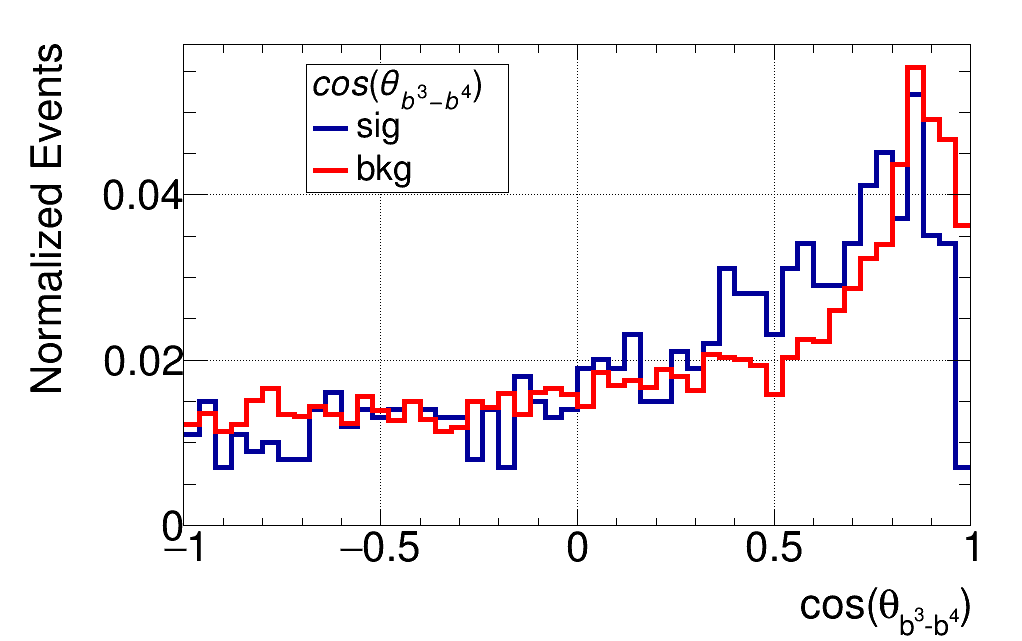}	
		\end{center}
	\end{minipage}
	\caption{The opening angle distributions between $b$-jet pairs emerging from $h^{1}$ (a) and  $h^{2}$ (b) for BP4 and background events are shown.}\label{f:bjet_angle}
\end{figure}

In relation to the two reconstructed  light Higgs bosons, we label $b^1,~b^2$ as the decay products of $h^1$ and  $b^3,~b^4$ as those of $h^2$. Thus, we can calculate the opening angle for the first two $b$-jets and the last two $b$-jets. Since they come from a very light Higgs boson for the signal, they will be highly boosted at the LHC, hence the two pairs of $b$-jets should have their opening angles close to zero. For the main $t\bar{t}$ background, if two normal jets emerge from the $W^\pm$ boson and  are mistagged as $b$-jets, there will be four $b$-jets and two of them (those coming from the $W^\pm$ boson) would also tend to be parallel but the other two (the true $b$-jets) would not. The distributions for the $b$-jet angles are shown in Fig.~\ref{f:bjet_angle} and confirm this picture.

\begin{figure}[ht]
	\begin{minipage}{0.47\textwidth}
		\begin{center} 
			(a)	\\
			\includegraphics[height=4.5cm]{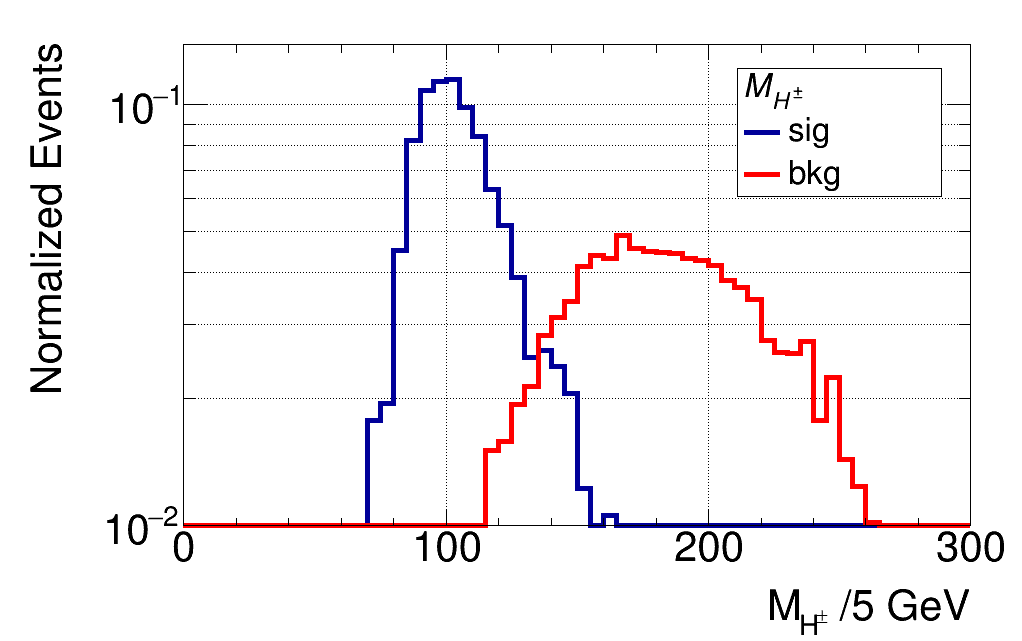}
			
		\end{center}
	\end{minipage}
	\begin{minipage}{0.47\textwidth}
		\begin{center} 
			(b)	\\
			\includegraphics[height=4.5cm]{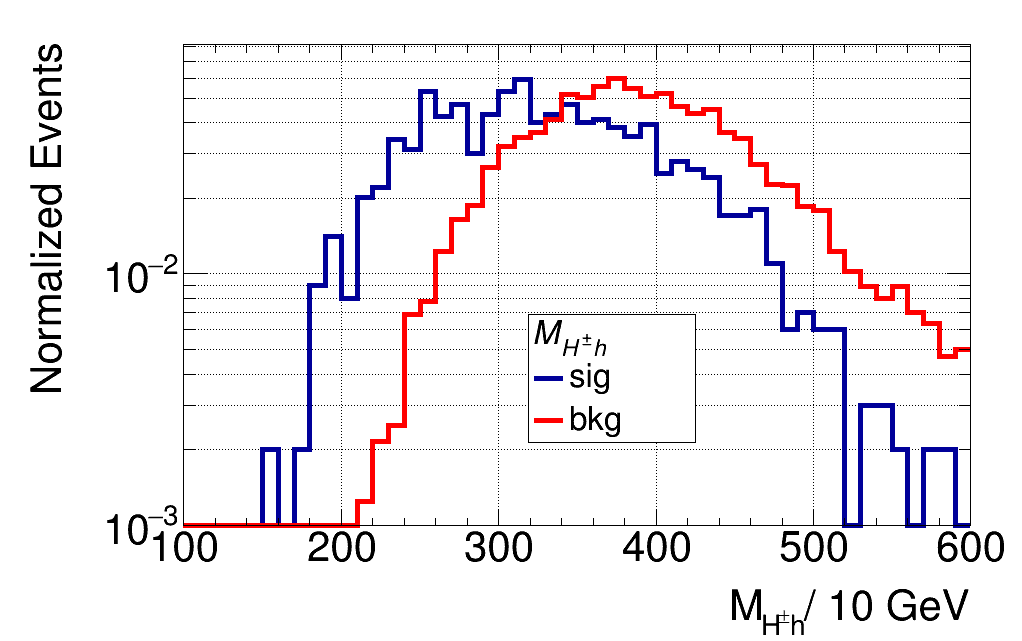}	
		\end{center}
	\end{minipage}
	\begin{minipage}{0.47\textwidth}
		\begin{center} 
			(c)	\\
			\includegraphics[height=4.5cm]{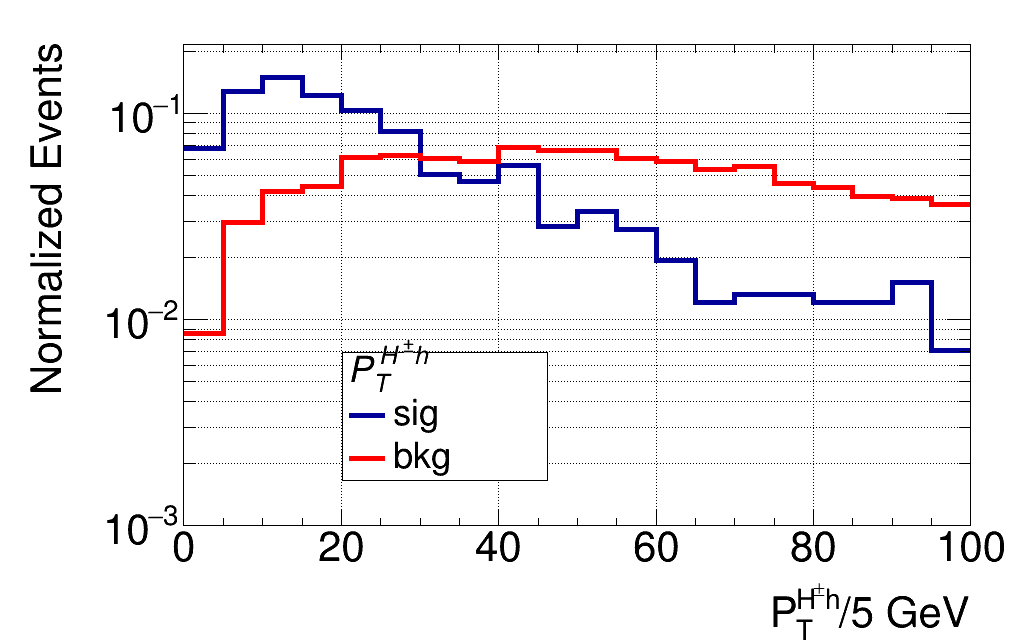}	
		\end{center}
	\end{minipage}
	\begin{minipage}{0.47\textwidth}
		\begin{center} 
			(d)	\\
			\includegraphics[height=4.5cm]{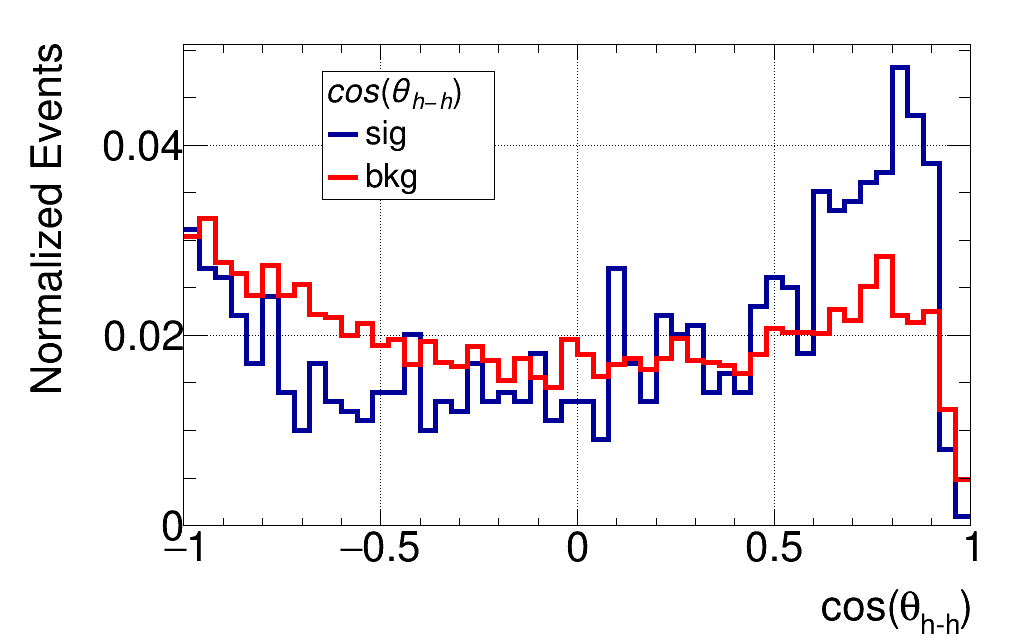}
			
		\end{center}
	\end{minipage}
	\caption{The reconstructed charged Higgs mass distributions (a), the invariant mass distributions (b) and transverse momentum distributions (c) of the neutral and charged Higgs boson system,  and the opening angle distributions of  the light Higgs pair (d) for BP4  and backgrounds events are shown.}\label{f:charged_higgs}
\end{figure}

Next, we need to reconstruct the charged Higgs boson mass. In doing this, we cannot first reconstruct the $W^\pm$ mass (like is done, e.g., in top (anti)quark searches~\cite{ATLAS:2012byx}) since this is off-shell. However,  we can use the same approach, directly applied to the $M_{H^\pm}$. Thus,   we  solve for the  neutrino longitudinal  momentum by using the lepton and light Higgs boson four-momentums alongside the MET. Since we have  reconstructed two light Higgs bosons, we use both of these to obtain the charged Higgs boson mass. Between the two ($h^1$ and $h^2$), we identify the one coming from the
$H^\pm \to W^{\pm *}h$ decay as the one that gives the best  charged Higgs mass (i.e., that  closer to the input value for $M_{H^\pm}$). The other light Higgs boson is then the state produced in association with the charged Higgs boson in $pp\to H^\pm h$. 
The mass distribution of  the correctly reconstructed charged Higgs boson  is shown in Fig.~\ref{f:charged_higgs}(a), wherein there is a clear difference between the signal and background events. The invariant mass distributions and the transverse momentum of  the  neutral and charged Higgs system, $H^\pm h$,  are also shown in Fig.~\ref{f:charged_higgs}(b), (c), also showing a difference between signal and background. In both cases, because  all resonances are light in the signal, the BSM peaks are much softer than the SM ones. In fact,  
 the opening angle distribution of the light Higgs boson pair for the signal is also different with respect to the background one since, as mentioned, the resonances in the signal are light enough to be boosted while this does not occur in the background, as seen in  Fig.~\ref{f:charged_higgs}(c).

\begin{figure}[ht]
\begin{minipage}{0.47\textwidth}
      \begin{center} 
      (a)\\
         \includegraphics[height=4.5cm]{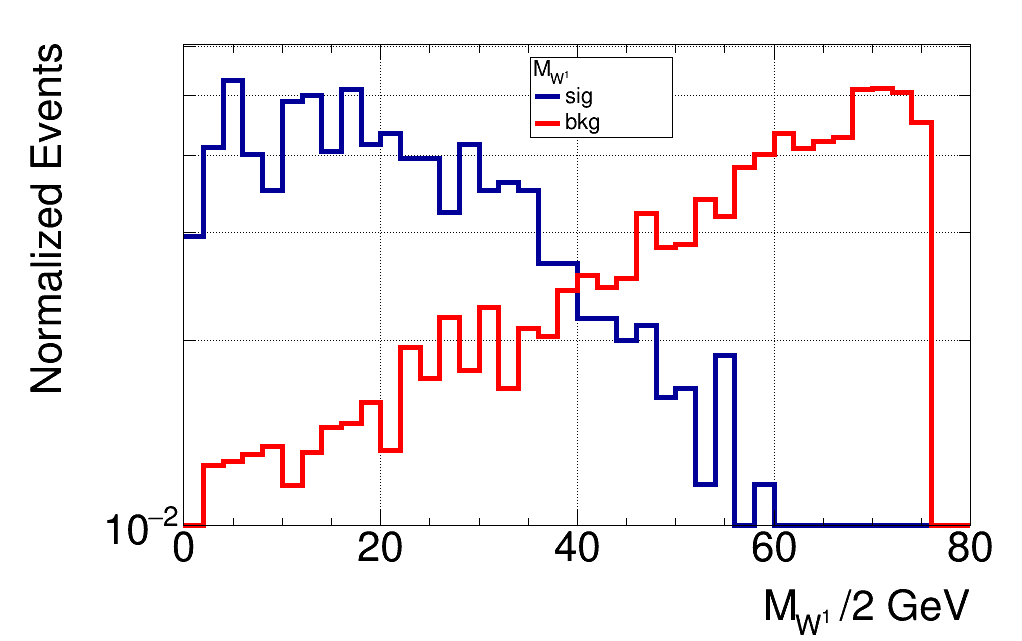}	
      \end{center}
\end{minipage}
\begin{minipage}{0.47\textwidth}
      \begin{center} 
      (b)\\
         \includegraphics[height=4.5cm]{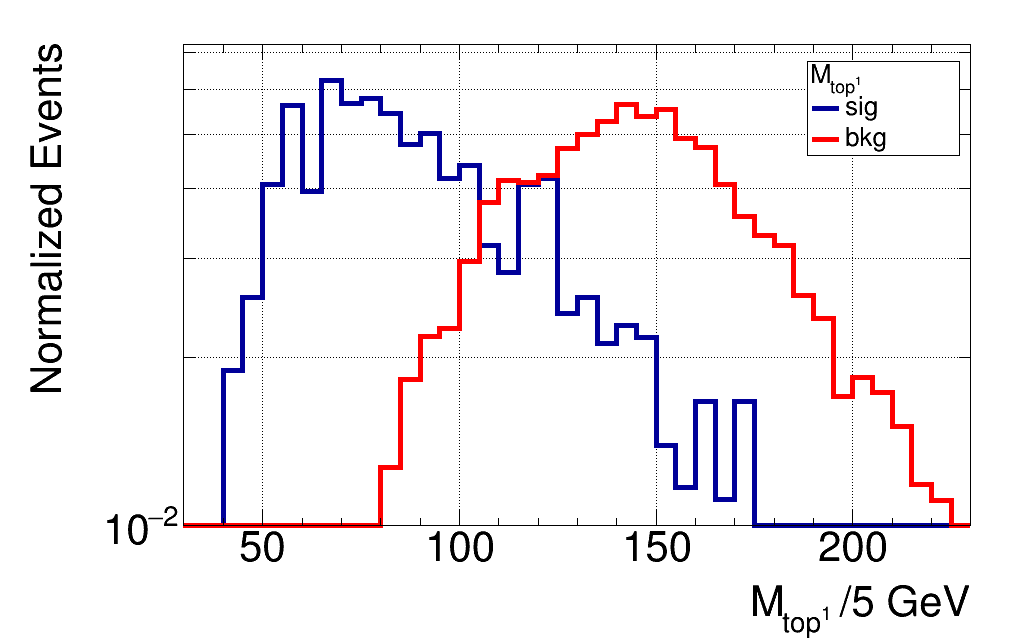}	
      \end{center}
\end{minipage}
\begin{minipage}{0.47\textwidth}
      \begin{center} 
            (c)\\
         \includegraphics[height=4.5cm]{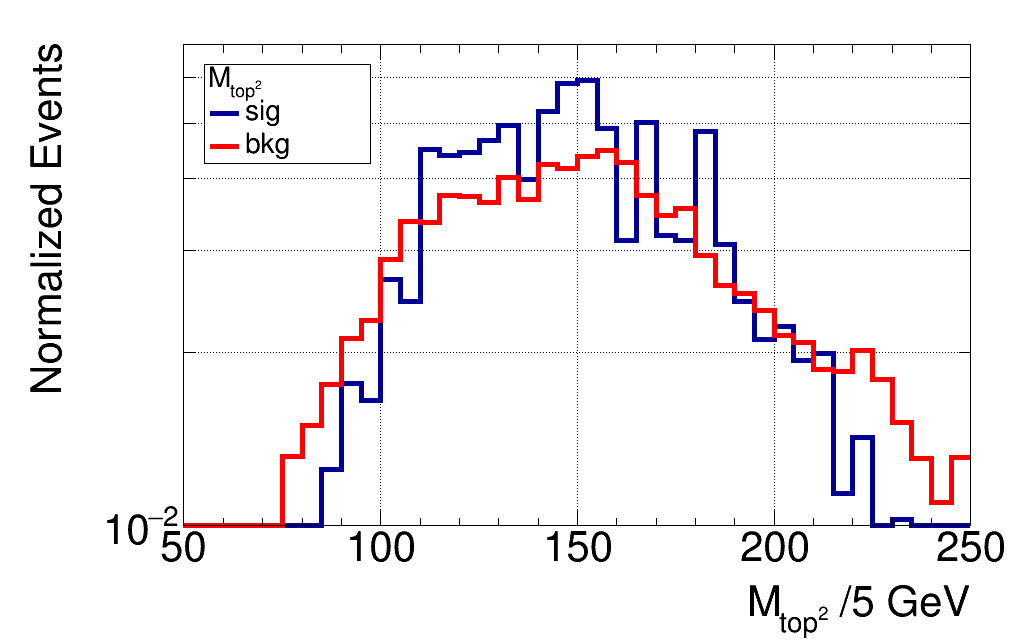}	
      \end{center}
\end{minipage}
\begin{minipage}{0.47\textwidth}
      \begin{center} 
            (d)\\
         \includegraphics[height=4.5cm]{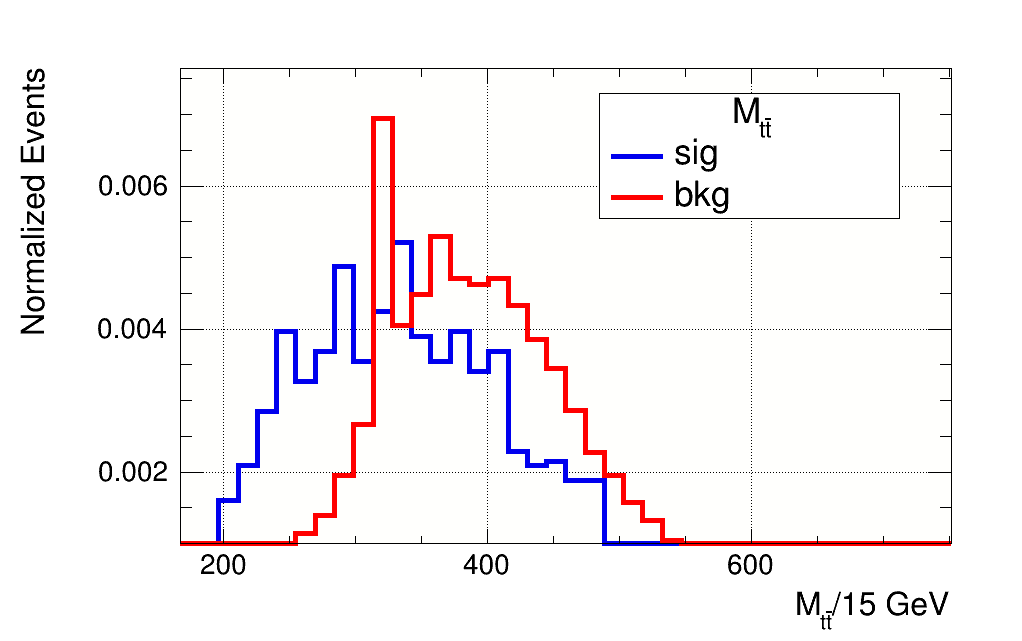}	
      \end{center}
\end{minipage}
\begin{minipage}{0.47\textwidth}
      \begin{center} 
                     (e)	\\
         \includegraphics[height=4.5cm]{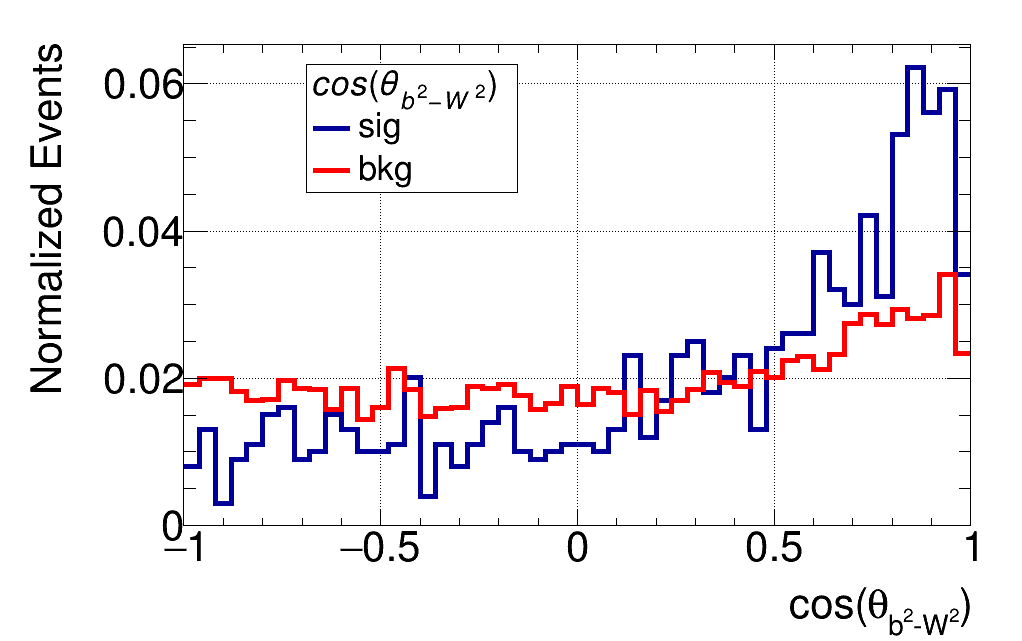}	
      \end{center}
\end{minipage}
 \caption{The reconstructed leptonic  $W^\pm$ boson  mass  distributions  (a), the reconstructed  leptonic (b) and hadronic (c)  top (anti)quark mass distributions, respectively, the $t\bar{t}$ system invariant mass  (d) and the angle between the $b$-jet from leptonic (anti)top quarks and the  hadronic $W^\pm$ boson  (e) for BP4 and background events are shown.}

\label{f:topinm}
\end{figure}

In order to significantly suppress the dominant background, i.e., $t\bar{t}$ production and decay via SM channels, it is necessary to veto such events. For this purpose, we reconstruct the two top (anti)quark masses. The lepton momentum and the MET are used to reconstruct the leptonically decaying $W^\pm$ boson first ($W^1$)~\cite{ATLAS:2012byx}. Because in the signal the $W^\pm$ boson is always off-shell, the peak will be much lighter here than the true $W^\pm$ boson mass but, for $t\bar{t}$ events, the $W^\pm$ boson is always on-shell, so a clear difference between the two distributions emerges, as shown in Fig.~\ref{f:topinm}(a). 
Then, we choose the softest of the two $b$-jets to reconstruct the other $W^\pm$ boson ($W^2$), noting that these two $b$-jets are mistagged  light quark/gluon-jets in the $t\bar{t}$ process. 
Next, we reconstruct the top (anti)quark pair with two hard $b$-jets and two reconstructed $W^\pm$ bosons. A $\chi^2$ is used to find the best combination of the leptonically ($M_{{\rm top}^{1}}$) and hadronically ($M_{{\rm top}^{2}}$) decaying top (anti)quarks, with testing function 
\begin{equation}
\chi^{2}=(M_{b^i W^1}-M_{{\rm top}^{1}})^2+(M_{b^j W^2}-M_{{\rm top}^{2}})^2,~~~i,j=1,2.
\end{equation}
For the $t\bar{t}$ background, the peak of the reconstructed top (anti)quark in leptonic mode will be around $m_t$ while the typical value of the peak will be much smaller in the signal, since the $W^\pm$ boson is off-shell. As for the hadronic top (anti)quark mass reconstruction, this is subject to more combinatorics, so it is
not expected to be extremely sharp in either case.  The  top (anti)quark mass distributions are shown in Fig.~\ref{f:topinm}(b) and (c), which demonstrate that our reconstruction procedure generally works well.
Based on two top (anti)quark reconstructions, we also plot the invariant mass of the $t\bar{t}$ system, which is shown in Fig.~\ref{f:topinm}(d). Finally, 
an angle, cos$(\theta_{b^{2}-W^{2}})$, is also shown: this is the opening angle between the $b$-jet in the leptonically decaying top (anti)quark and the $W^\pm$ boson in the hadronically decaying top (anti)quark. For the $t\bar{t}$ process, there will be no apparent  tendency. However, in the signal process, because both the charged Higgs boson and the neutral Higgs boson are light, their decay products will be boosted, thereby ending up  parallel to each other. In other words, the angle of three $b$-jets will be small, no matter which three $b$-jets are chosen. This angle distribution is shown in Fig.~\ref{f:topinm}(e).

\subsection{The TMVA inputs and results} 
To improve and optimize the signal and background distinction, we use the  Gradient-Boosted Decision Tree (GBDT) approach, which is implemented in the Toolkit for Multi-Variant Analysis (TMVA) within Root~\cite{Therhaag:2010zz}. 
We first use a very loose kinematic selection before the TMVA training for data clean, which only contains the cuts for transverse momentum and pseudo-rapidity as well as very loose $M_{4b}$, $M_{b\bar{b}}^{1}$, $M_{b\bar{b}}^{2}$, $M_{H^{\pm}}$, $         P^{T}_{H^{\pm}h} $, $M_{W^1}$, $M_{{\rm top}^1}$ and $M_{{\rm top}^2}$ cuts,
so as to make  sure that input data are not polluted by outliers.  The loose kinematic cuts are shown in the first column in Tab.~\ref{t:selection_cuts}. 
 
 \begin{table}
 \begin{center}
\begin{tabular}{|c| c|  c|}
 \hline
          &        Loose kinematic cuts          & Tight kinematic cuts\\
 \hline
 $                M_{4b}  $ & $ [ 100 ,600 ] $  & $[ 100 ,450 ] $\\
 \hline
 $              M_{bb}^{1}$ & $  [ 10 ,110 ] $ &$[ 10 ,75 ] $\\
 \hline
  $              M_{bb}^{2}$ & $               [ 10 ,150 ] $ & $[ 25 ,100 ] $\\
 \hline
 $             M_{H^{\pm}} $ & $     [ 50 ,250 ] $  & $[ 60 ,190 ] $\\
  \hline
 $         P^{T}_{H^{\pm}h} $ & $   [ 0 ,110 ] $  & $[ 0 ,90 ] $\\
  \hline
 $                M_{W^{1}} $ & $             [ 0 ,100 ] $  & $[ 0 ,80 ] $\\
  \hline
  $             M_{{\rm top}^{1}} $ &  $  [ 20 ,230 ] $ & $[ 30 ,230 ] $\\
   \hline
  $             M_{{\rm top}^{2}}  $& $   [ 50 ,280 ] $ & $[ 70 ,280 ] $\\
 \hline
 MVA cut                            &     -             & $[ 0.5, 1] $\\
  \hline
 \end{tabular}
 \end{center}
 \caption{ The pre-selection loose and tight kinematic cuts for BP4 and background events are shown. Except in the last line, all numerical values are in GeV.}\label{t:selection_cuts}
  \end{table}

\begin{figure}[ht]
\begin{minipage}{0.47\textwidth}
      \begin{center} 
                     (a)	\\
         \includegraphics[height=4.5cm]{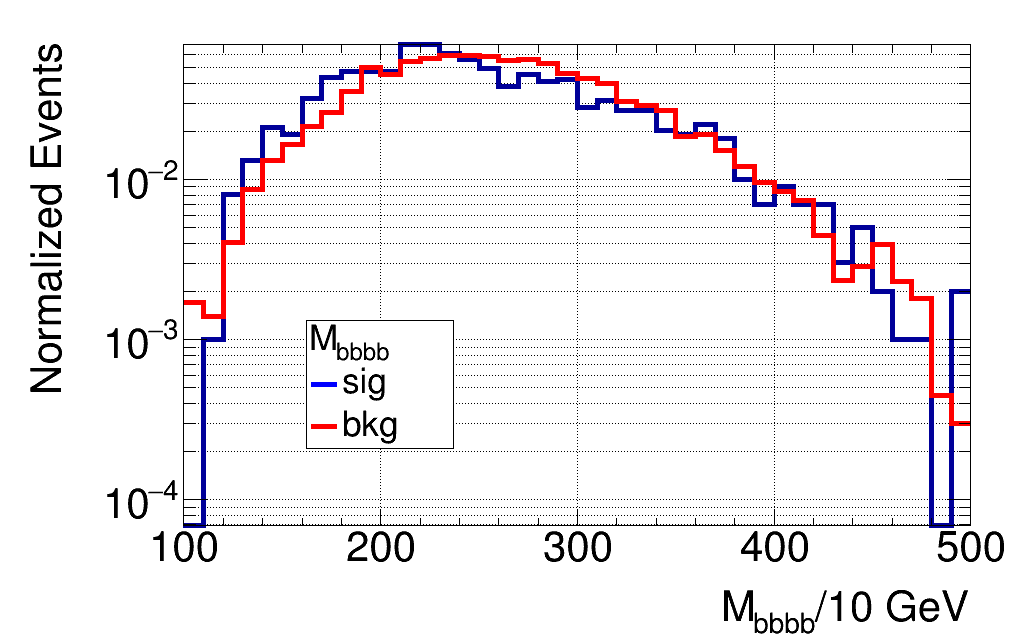}	
      \end{center}
\end{minipage}
\begin{minipage}{0.47\textwidth}
      \begin{center} 
                     (b)	\\
         \includegraphics[height=4.5cm]{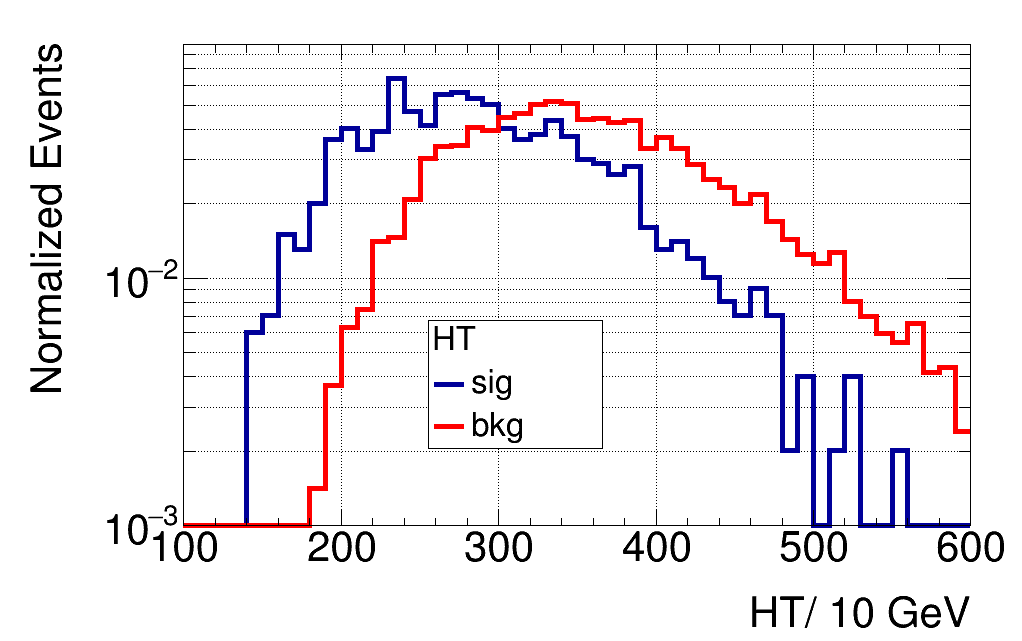}	
      \end{center}
\end{minipage}
 \caption{The invariant mass of the  four $b$-jets (a) and the scalar sum of the transverse momenta of visible particles  for BP4 and  background events are shown.}\label{f:visible}
\end{figure}

In the training stage, we used 13 input variables in total for the TMVA, which are shown in Tab.~\ref{t:MVA_observables}. These observables are divided into three categories. 
The first category is related to the possible resonances in the signal while the second is made up of the variables characterizing the $t\bar{t}$ background, all of which have been described above. The third kind uses generic final state variables, like the invariant mass of the four $b$-jets and the scalar sum of the visible particle transverse momenta ($HT$), both of which are shown in  Fig.~\ref{f:visible}.

\begin{table}[H]
 \begin{center}
 \begin{small}
\centerline{}
\begin{tabular}{|c|c|c|c|c|}
\hline
 BSM invariant masses& $M_{b\bar{b}}^{1}$& $M_{b\bar{b}}^{2}$ &$M_{H^{\pm}}$ & $M_{H^\pm}-M_h$\\
\hline
BSM angles &$\cos(\theta_{b^{1}-b^{2}})$ & $\cos(\theta_{b^{3}-b^{4}})$ & $\cos(\theta_{h-h})$ &\\
\hline
\hline
SM invariant masses & $M_{W^1}$ & $M_{{\rm top}^1}$ & $M_{t\bar t}$& \\
\hline
SM angles &$\cos(\theta_{b^{2}-W^{2}})$& &&\\
\hline
\hline
Other variables & $M_{4b}$ & $HT$   &    &\\

\hline
 
\end{tabular}
 \end{small}
   \caption{The input observables used in the TMVA analysis using a GBDT.}\label{t:MVA_observables} 
 \end{center}
  \end{table}

\begin{figure}[ht]
      \begin{center}   
         \includegraphics[height=4.5cm]{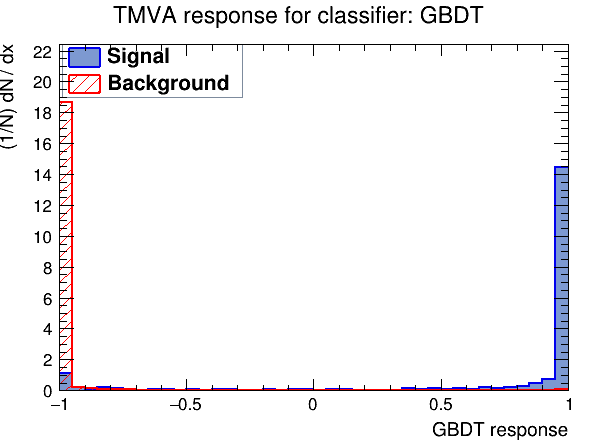}	
      \end{center}
 \caption{The TMVA response for classifier GBDT of  BP4 and  background events.}\label{f:MVA}
\end{figure}

Generally speaking, it is found that  the invariant mass related observables of signal and background events are more powerful than the angle related ones. Anyhow, all are used and we apply the GBDT model on the signal and background to calculate their final scores, after which there is a very clear separation between the two, which is shown in Fig.~\ref{f:MVA}. Finally, we apply the tight kinematic cuts from the second column in Tab.~\ref{t:selection_cuts} to extract the signal significances.

\subsection{Significances at  LHC Run 3}

After all the described cuts, the significances for each (multiplicity) category of the final state are found and we have summarised these in Tab.~\ref{t:sig_bp20}. 
%
Here, a few comments are in order. For LACs, most of the BPs can have a large significance for all three categories. For TACs, most of the significances can be larger than 3 when the final state is the 4b0j case. Further, for all BPs, we can achieve a large enough significance, above and beyond discovery in all cases,  by combining the three categories of signatures. From these BPs, it is also observed that the significances mainly depend on the signal cross section and the charged Higgs boson mass. 
In fact, it is obvious  that a larger charged Higgs mass will generate a harder lepton and $b$-jets, which will in turn  increase the reconstruction efficiencies for these objects in the pursued final state.

 \begin{table}
  \begin{center}
 \begin{small}
\begin{tabular}{|c|c|c|c|c|c|c|} 
\hline
   & \multicolumn{3}{c|}{LACs}& \multicolumn{3}{c|}{TACs}\\
\hline
         &  2b2j & 3b1j & 4b0j &  2b2j & 3b1j & 4b0j\\
\hline
BP1 & 3.65 & 8.51 & 8.79 &0.45 & 1.60 & 3.28 \\
\hline
BP2 & 2.19 & 5.10 & 6.06 & 0.27 & 1.30 & 2.45\\
 \hline
 BP3 & 3.01 & 6.82 & 7.21 &0.51 & 1.90 & 3.3\\
\hline
BP4 & 3.56  & 8.12  & 9.08 & 0.73 & 2.97  & 5.44\\
\hline
 BP5 & 3.55  & 7.96  & 9.43 & 0.71 & 2.42  & 4.91\\
\hline
BP6 & 2.85  & 6.41  & 7.74 & 0.70 & 2.37  & 4.79\\
\hline
\end{tabular}
 \end{small}
  \caption{The significances for our BPs with both LACs and TACs are shown. Rates are for $\sqrt{s}$ = 14 TeV and $L = 300~\rm{fb}^{-1}$.}\label{t:sig_bp20}
 \end{center}
  \end{table}

\begin{figure}[ht]
	\begin{minipage}{0.47\textwidth}
		\begin{center} 
			\includegraphics[height=4.5cm]{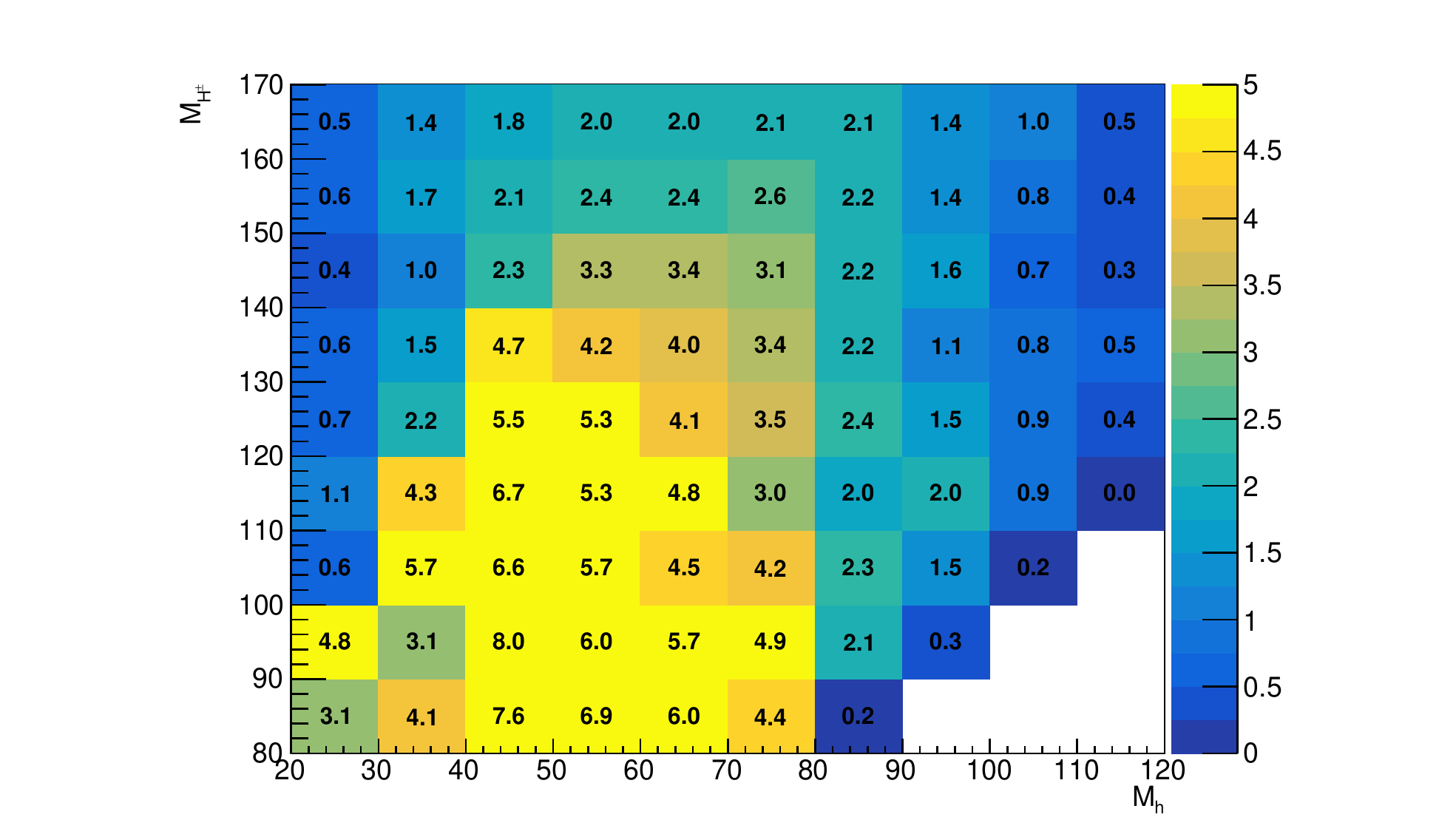}	
		\end{center}
	\end{minipage}
	\begin{minipage}{0.47\textwidth}
		\begin{center} 
			\includegraphics[height=4.5cm]{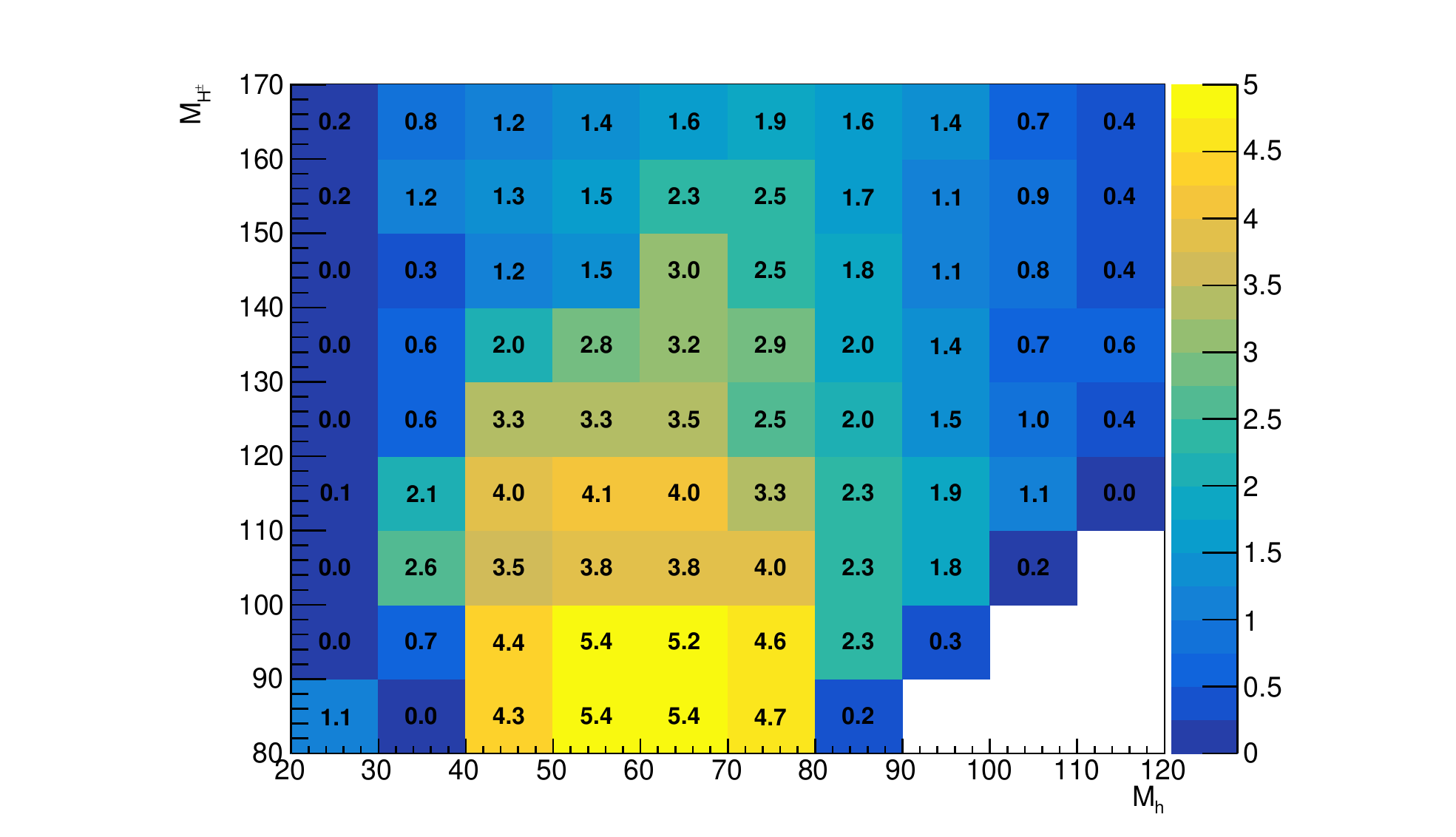}	
		\end{center}
	\end{minipage}
	\caption{The predicted significances over the ($M_{h}$, $M_{H^{\pm}}$) plane for the 4b0j case with both LACs and TACs are shown,  where $\sqrt{s}$ = 14 TeV and $L = 300~\rm{fb}^{-1}$.}\label{f:hot_map_mass}
\end{figure}

\begin{figure}[ht]
\begin{minipage}{0.47\textwidth}
      \begin{center} 
         \includegraphics[height=4.5cm]{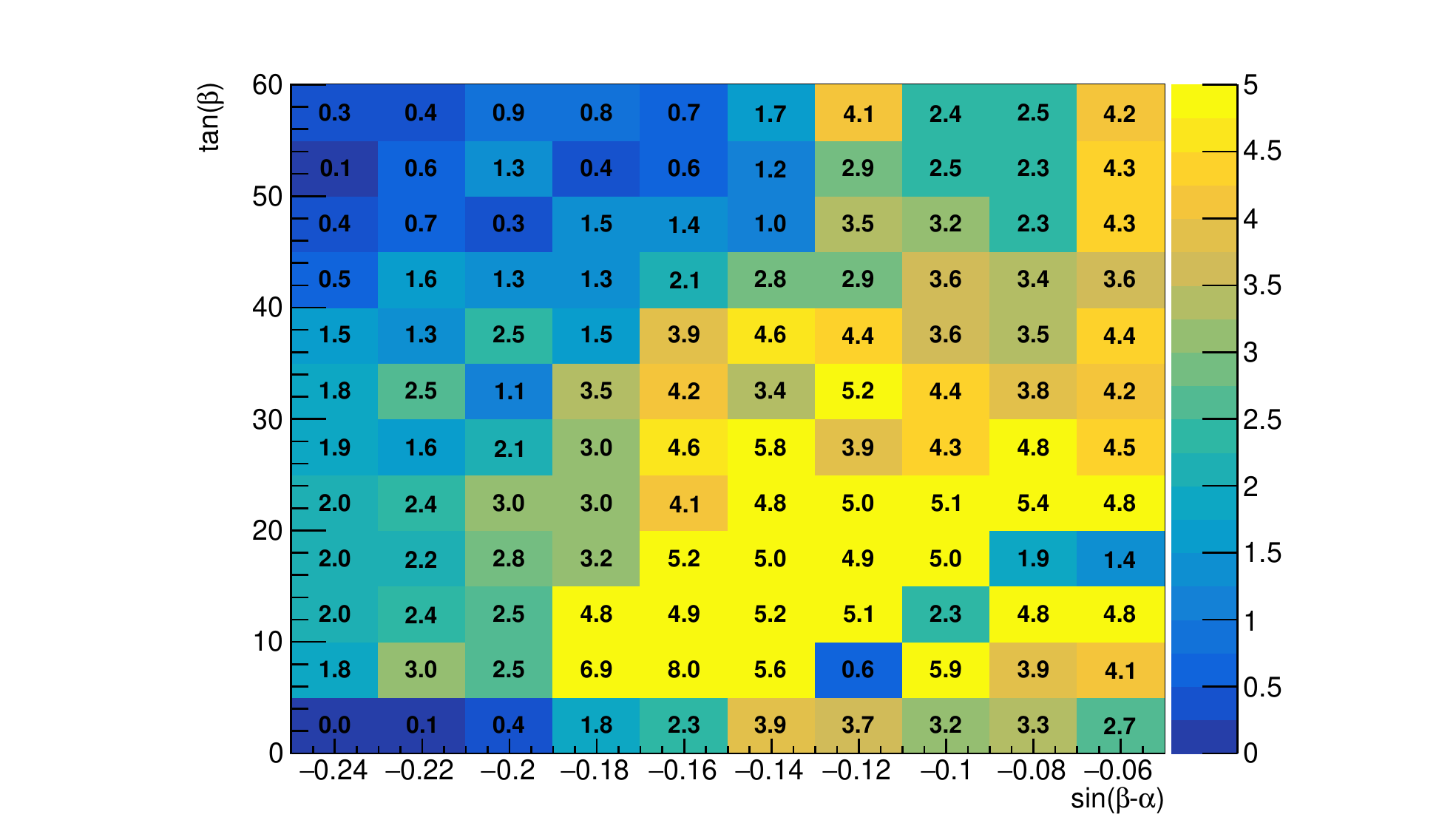}	
      \end{center}
\end{minipage}
\begin{minipage}{0.47\textwidth}
      \begin{center} 
         \includegraphics[height=4.5cm]{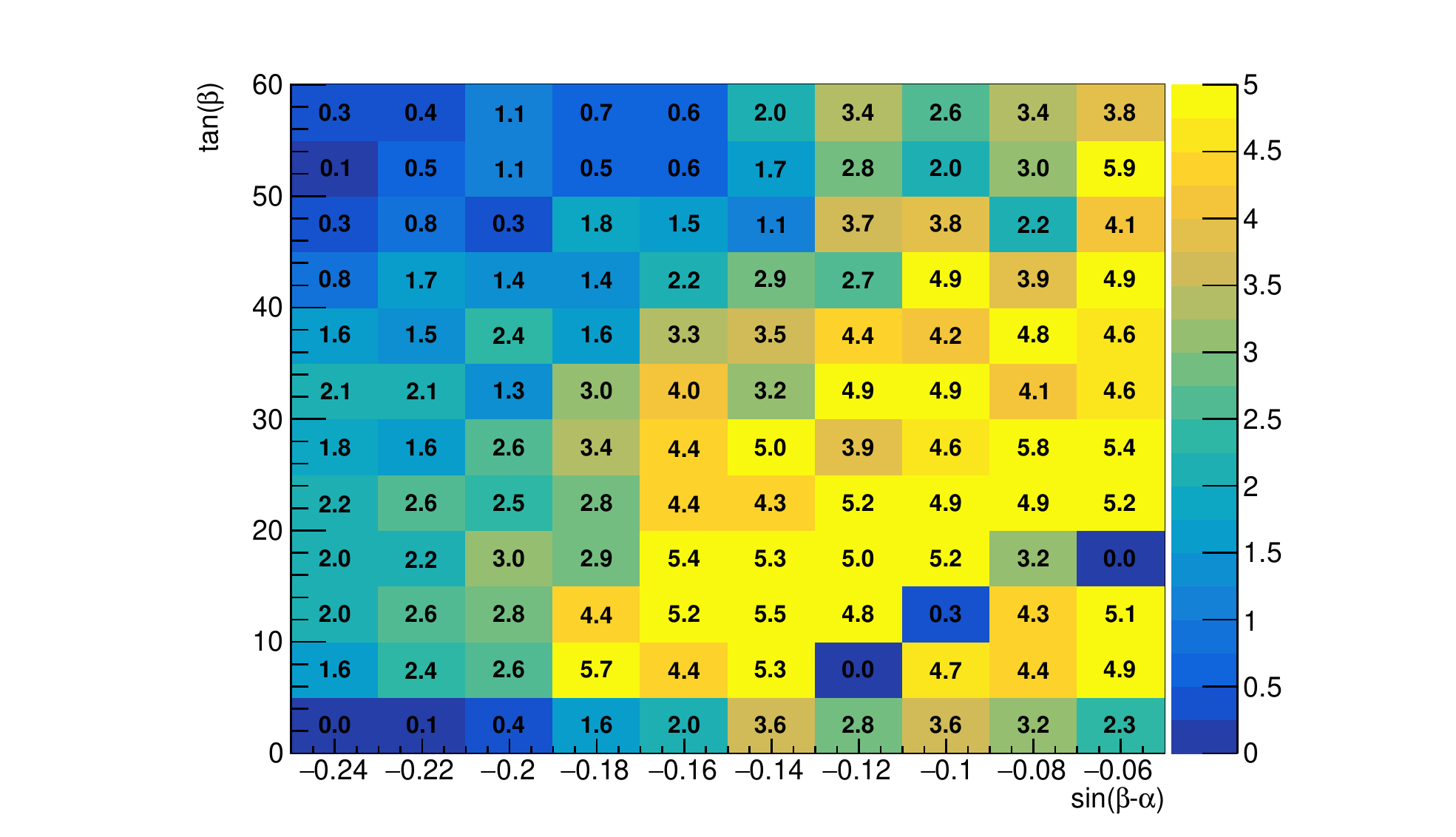}	
      \end{center}
\end{minipage}
 \caption{The predicted significances over the ($\sin(\beta-\alpha)$, $\tan\beta$) plane for the 4b0j case with both LACs and TACs are shown, where $\sqrt{s}$ = 14 TeV and $L = 300~\rm{fb}^{-1}$.}\label{f:hot_map_para}
\end{figure}

In order to have a panoramic view of the model parameter space, we take the 4b0j case as an example and explore the feasibility of the LHC when $\sqrt{s}$ = 14 TeV and $L = 300~\rm{fb}^{-1}$. The significances for the model parameter space are exposed in the heatmaps of Figs.~\ref{f:hot_map_mass} and \ref{f:hot_map_para}. 

The significances over the ($M_{h}$, $M_{H^{\pm}}$) plane are shown in Fig.~\ref{f:hot_map_mass}. To obtain the results given here, the ($M_{h}$, $M_{H^{\pm}}$) plane is divided into 90 grids, with $M_{h}$ in $(20,120) $ GeV and $M_{H^{\pm}}$ in $(80,170) $ GeV.
In each grid,  the free parameters $\tan\beta$ and $\sin(\beta-\alpha)$ are scanned for all possible allowed values and the BPs which have the maximal theoretical cross sections are taken for this grid. The   events are generated with both  LACs and TACs  and the significances are calculated in each grid as in the above section. 
From this figure, we notice that there are clear signals when $M_h $ is  in $ (40-80) $ GeV and $M_{H^{\pm}} $ is  in $  (80-130) $ GeV for both LACs and TACs. The maximum significance could reach 8.0 and 5.4 for LACs and TACs, respectively.
 A similar heatmap is also made over  the ($\sin(\beta-\alpha)$, $\tan\beta$) plane, as shown in Fig.~\ref{f:hot_map_para}. The sensitivity regions for these two parameters are found as $-0.18 < \sin(\beta-\alpha)< -0.04$ and $5< \tan\beta < 40$, respectively.
 
Finally, it should be pointed out that these two heatmaps, for the ($M_{h}$, $M_{H^{\pm}}$) and ($\sin(\beta-\alpha)$, $\tan\beta$) planes, are consistent with each other. For LACs and TACs, the maximum significances are the same within the error bars. The slight difference can be attributed to the statistic uncertainties in generating MC events.

\section{Conclusions}\label{sum}
In this paper, we have  performed a detailed analysis of the $pp \rightarrow H^{\pm} h \rightarrow W^{\pm*} h h \rightarrow \ell^\pm \nu+ 4b$ ($\ell=e,\mu$) process in the 2HDM Type-I at the LHC.  By simulating the full event chain, including hard scattering,  parton shower, hadronization and detector effects using 
mainstream numerical tools, we have obtained realistic predictions for both signal and background processes. This has enabled us to assess the feasibility of observing the signal in a realistic experimental environment.

To optimize sensitivity to the signal while suppressing the background contributions, initially, we have carefully chosen kinematic cuts in pseudorapidity, transverse momentum, cone separation and MET. These cuts, represented by two sets of kinematic selections, one loose and one tight, have ensured an efficient event selection while maintaining a reasonable signal-to-background ratio. 
Then, we have categorized events  based on the number of $b$- as well as light quark/gluon-jets present, resulting in three distinct event categories. This categorization has allowed us to investigate the specific signatures associated with each jet multiplicity and design dedicated analysis strategies for enhanced signal extraction, leveraging the kinematic features (chiefly, resonant masses) of
both signal and backgrounds, again, in the presence of a loose and tight selection. This has proven successful, as we have been able to establish sensitivity to the described signal already by the end of Run 3 of the LHC. We have demonstrated this by  presenting heatmaps in the ($M_{h}$, $M_{H^{\pm}}$)  and ($\sin(\beta-\alpha)$, $\tan\beta$) planes to visualize  regions of parameter space with enhanced signal significance. Finally, 
in order to facilitate experimental analyses and provide guidance for future phenomenological studies, we have presented six BPs, each being  carefully selected to cover a region of parameter space exhibiting interesting spectrum properties, such as relatively light charged and neutral Higgs bosons  and an off-shell $ W^{\pm *}$.

In summary, our study provides a comprehensive analysis of a hallmark  process of the 2HDM Type-I at the LHC, which may enable one to verify the BSM nature of the EWSB mechanism.

\acknowledgments
The work of AA, RB, MK and BM is supported by the Moroccan Ministry of Higher
Education and Scientific Research MESRSFC and CNRST Project PPR/2015/6. The
work of SM is supported in part through the NExT Institute and STFC Consolidated
Grant No. ST/L000296/1. ZL's work is supported by the Graduated Research and Innovation Fund Project of Inner Mongolia Normal University No. CXJJS21129. YW's work is supported by the Natural Science Foundation of
China Grant No. 12275143, the Inner Mongolia Science Foundation  Grant No. 2020BS01013 and Fundamental Research Funds for the Inner Mongolia Normal University Grant No. 2022JBQN080. QSY’s work is supported by the Natural Science Foundation of
China Grant No. 12275143 and No. 11875260.


\begin{thebibliography}{100}	
	
	\bibitem{Aad:2012tfa}
	G.~Aad \textit{et al.} [ATLAS],
	Phys. Lett. B \textbf{716} (2012)  1,
	[arXiv:1207.7214 [hep-ex]].
	
	\bibitem{Chatrchyan:2012ufa}
	S.~Chatrchyan \textit{et al.} [CMS],
	Phys. Lett. B \textbf{716} (2012)  30,
	[arXiv:1207.7235 [hep-ex]].
		
	\bibitem{Lee:1973iz}
	T.~D.~Lee,
	Phys. Rev. D \textbf{8} (1973), 1226-1239.
	
	\bibitem{Deshpande:1977rw}
	N.~G.~Deshpande and E.~Ma,
	Phys. Rev. D \textbf{18} (1978), 2574.
	
	\bibitem{Branco:2011iw}
	G.~C.~Branco, P.~M.~Ferreira, L.~Lavoura, M.~N.~Rebelo, M.~Sher and J.~P.~Silva,
	Phys. Rept. \textbf{516} (2012), 1-102
	[arXiv:1106.0034 [hep-ph]].
	
	\bibitem{Glashow:1976nt}
	S.~L.~Glashow and S.~Weinberg,
	Phys. Rev. D \textbf{15} (1977), 1958
	
	\bibitem{Akeroyd:2016ymd}
	A.~G.~Akeroyd, M.~Aoki, A.~Arhrib, L.~Basso, I.~F.~Ginzburg, R.~Guedes, J.~Hernandez-Sanchez, K.~Huitu, T.~Hurth and M.~Kadastik, \textit{et al.}
	Eur. Phys. J. C \textbf{77} (2017), no.5, 276
	[arXiv:1607.01320 [hep-ph]].	
	
	\bibitem{Arhrib:2021xmc}
	A.~Arhrib, R.~Benbrik, M.~Krab, B.~Manaut, S.~Moretti, Y.~Wang and Q.~S.~Yan,
	JHEP \textbf{10} (2021), 073
	[arXiv:2106.13656 [hep-ph]].
	
	\bibitem{Akeroyd:1998dt}
	A.~G.~Akeroyd,
	Nucl. Phys. B \textbf{544} (1999), 557-575
	[arXiv:hep-ph/9806337 [hep-ph]].
	
	\bibitem{Arhrib:2016wpw}
	A.~Arhrib, R.~Benbrik and S.~Moretti,
	Eur. Phys. J. C \textbf{77} (2017), no.9, 621 
	[arXiv:1607.02402 [hep-ph]].
	
	\bibitem{Bahl:2021str}
	H.~Bahl, T.~Stefaniak and J.~Wittbrodt,
	JHEP \textbf{06} (2021), 183
	[arXiv:2103.07484 [hep-ph]].
	
	\bibitem{Cheung:2022ndq}
	K.~Cheung, A.~Jueid, J.~Kim, S.~Lee, C.~T.~Lu and J.~Song,
	Phys. Rev. D \textbf{105} (2022) no.9, 095044
	[arXiv:2201.06890 [hep-ph]].
	
	\bibitem{Mondal:2023wib}
	T.~Mondal, S.~Moretti, S.~Munir and P.~Sanyal,
	[arXiv:2304.07719 [hep-ph]].

	
\bibitem{Bhatia:2022ugu}
D.~Bhatia, N.~Desai and S.~Dwivedi,
[arXiv:2212.14363 [hep-ph]].


\bibitem{Bandyopadhyay:2015dio}
P.~Bandyopadhyay, K.~Huitu and S.~Niyogi,
JHEP \textbf{07}, 015 (2016)
doi:10.1007/JHEP07(2016)015
[arXiv:1512.09241 [hep-ph]].
	
	\bibitem{Sanyal:2023pfs}
	P.~Sanyal and D.~Wang,
	[arXiv:2305.00659 [hep-ph]].
		
	\bibitem{Arhrib:2021yqf}
	A.~Arhrib, R.~Benbrik, M.~Krab, B.~Manaut, S.~Moretti, Y.~Wang and Q.~S.~Yan,
	Symmetry \textbf{13} (2021) no.12, 2319
	[arXiv:2110.04823 [hep-ph]].
	
	\bibitem{Wang:2021pxc}
	Y.~Wang, A.~Arhrib, R.~Benbrik, M.~Krab, B.~Manaut, S.~Moretti and Q.~S.~Yan,
	JHEP \textbf{12} (2021), 021
	[arXiv:2107.01451 [hep-ph]].
	
	\bibitem{Wang:2021zjp}
	Y.~Wang, A.~Arhrib, R.~Benbrik, M.~Krab, B.~Manaut, S.~Moretti and Q.~S.~Yan,
	[arXiv:2111.12286 [hep-ph]].
	
	\bibitem{Eriksson:2009ws}
	D.~Eriksson, J.~Rathsman and O.~St{\aa}l,
	Comput. Phys. Commun. \textbf{181} (2010), 189-205
	[arXiv:0902.0851 [hep-ph]].
		
	\bibitem{Chang:2015goa}
	S.~Chang, S.~K.~Kang, J.~P.~Lee and J.~Song,
	Phys. Rev. D \textbf{92} (2015) no.7, 075023
	[arXiv:1507.03618 [hep-ph]].
	
	\bibitem{Kanemura:1993hm}
	S.~Kanemura, T.~Kubota and E.~Takasugi,
	Phys. Lett. B \textbf{313} (1993), 155-160
	[arXiv:hep-ph/9303263 [hep-ph]].
	
	\bibitem{Akeroyd:2000wc}
	A.~G.~Akeroyd, A.~Arhrib and E.~M.~Naimi,
	Phys. Lett. B \textbf{490} (2000), 119-124
	[arXiv:hep-ph/0006035 [hep-ph]]. A.~Arhrib,
	[arXiv:hep-ph/0012353 [hep-ph]].
	
	\bibitem{Peskin:1990zt}
	M.~E.~Peskin and T.~Takeuchi,
	Phys. Rev. Lett. \textbf{65} (1990), 964-967.
	
	\bibitem{Peskin:1991sw}
	M.~E.~Peskin and T.~Takeuchi,
	Phys. Rev. D \textbf{46} (1992), 381-409.
	
	\bibitem{Haller:2018nnx}
	J.~Haller, A.~Hoecker, R.~Kogler, K.~M\"onig, T.~Peiffer and J.~Stelzer,
	Eur. Phys. J. C \textbf{78} (2018), no.8, 675
	[arXiv:1803.01853 [hep-ph]].
	
	\bibitem{Bechtle:2020pkv}
	P.~Bechtle, D.~Dercks, S.~Heinemeyer, T.~Klingl, T.~Stefaniak, G.~Weiglein and J.~Wittbrodt,
	Eur. Phys. J. C \textbf{80} (2020) no.12, 1211
	[arXiv:2006.06007 [hep-ph]].
	
	\bibitem{Bechtle:2020uwn}
	P.~Bechtle, S.~Heinemeyer, T.~Klingl, T.~Stefaniak, G.~Weiglein and J.~Wittbrodt,
	Eur. Phys. J. C \textbf{81} (2021) no.2, 145
	[arXiv:2012.09197 [hep-ph]].
	
	\bibitem{HFLAV:2016hnz}
	Y.~Amhis \textit{et al.} [HFLAV],
	Eur. Phys. J. C \textbf{77} (2017) no.12, 895
	doi:10.1140/epjc/s10052-017-5058-4
	[arXiv:1612.07233 [hep-ex]].
	
	\bibitem{LHCb:2017rmj}
	R.~Aaij \textit{et al.} [LHCb],
	Phys. Rev. Lett. \textbf{118} (2017) no.19, 191801
	[arXiv:1703.05747 [hep-ex]].
	
	\bibitem{Mahmoudi:2008tp}
	F.~Mahmoudi,
	Comput. Phys. Commun. \textbf{180} (2009), 1579-1613
	[arXiv:0808.3144 [hep-ph]].
	
	\bibitem{Alwall:2014hca}
	J.~Alwall, R.~Frederix, S.~Frixione, V.~Hirschi, F.~Maltoni, O.~Mattelaer, H.~S.~Shao, T.~Stelzer, P.~Torrielli and M.~Zaro,
	JHEP \textbf{07} (2014), 079
	[arXiv:1405.0301 [hep-ph]].
	
	\bibitem{Sjostrand:2006za}
	T.~Sjostrand, S.~Mrenna and P.~Z.~Skands,
	JHEP \textbf{05} (2006), 026
	[arXiv:hep-ph/0603175 [hep-ph]].
	
	\bibitem{Sjostrand:2014zea}
	T.~Sj\"ostrand, S.~Ask, J.~R.~Christiansen, R.~Corke, N.~Desai, P.~Ilten, S.~Mrenna, S.~Prestel, C.~O.~Rasmussen and P.~Z.~Skands,
	Comput. Phys. Commun. \textbf{191} (2015), 159-177
	[arXiv:1410.3012 [hep-ph]].
	
	\bibitem{deFavereau:2013fsa}
	J.~de Favereau \textit{et al.} [DELPHES 3],
	JHEP \textbf{02} (2014), 057
	[arXiv:1307.6346 [hep-ex]].
	
\bibitem{ATLAS:2012byx}
G.~Aad \textit{et al.} [ATLAS],
Phys. Lett. B \textbf{717} (2012), 330-350
[arXiv:1205.3130 [hep-ex]].


\bibitem{Therhaag:2010zz}
J.~Therhaag,
PoS \textbf{ICHEP2010} (2010), 510




\end{thebibliography}
\end{document}